# Thermodynamic and Transport Properties Modeling of Deep Eutectic Solvents: A review on $g^E$-models, equations of state and molecular dynamics

*Andrés González de Castilla, Jan Philipp Bittner, Simon Müller, Sven Jakobtorweihen, Irina Smirnova\**

Institute of Thermal Separation Processes, Hamburg University of Technology, Eißendorfer Straße 38 (O) 21073, Hamburg. Germany

**Abstract**

Deep eutectic solvents (DESs) have gained attention in recent years as attractive alternatives to traditional solvents. There is a growing number of publications dealing with the thermodynamic modeling of DESs highlighting the importance of modeling the solutions' properties. In this review, we summarize the state-of-the-art in DES modeling as well as its current challenges. We also summarize the various modeling approaches to phase equilibria and properties of DESs with $g^E$-models, EOS and molecular dynamics (MD) simulations. The current $g^E$-model and EOS-based approaches handle DESs as pseudo-components in order to simplify the parameterizations and calculation strategies. However, for the models to become more transferable and predictive, it would be preferable to model the individual DES constituents instead of using the pseudo-components. This implies that validation with more detailed experimental data that includes the distribution of the DES components is also required. MD simulations, in contrast to $g^E$-models and EOS, are capable of providing information about the liquid structure and can predict dynamic properties although, the latter quantities still show some imprecisions. Therefore, insights into the liquid structure of DES systems from MD could also aid in improving present modeling strategies in addition to a better understanding. Finally, the latest developments for DES force fields are discussed as the quality of the applied force fields determine the results of MD simulations.

Deep eutectic solvents, $g^E$-model, equation of state, phase equilibria, molecular dynamics, force field

1. Introduction

Green chemistry and sustainability have become topics of growing interest in state of the art chemical engineering research and applications. In the search for economic solvents that have better environmental and operational properties, deep eutectic solvents (DESs) have gained attention as an attractive alternative to traditional solvents.[1–4] DESs, first described in 2003 by Abbott et al.[5], are formed by mixing different molar ratios of a polyatomic cation with a Lewis base as counter-ion and a hydrogen bond donor (HBD: either a Lewis or a Brønsted acid). The salt acts as a hydrogen bond acceptor (HBA) and the hydrogen bonding interactions in these mixtures lead to melting points that are considerably lower than those of the salt and the HBD[6]. Non-ionic DESs are also described in the literature.[1,7] The central characteristic of a DES is that it constitutes a liquid HBA:HBD mixture with an eutectic distance[1] that, as recently proposed by Martins et al.[8], increases due to strong negative deviations from ideal behavior. This is depicted for choline chloride (ChCl) + urea in Figure 1.

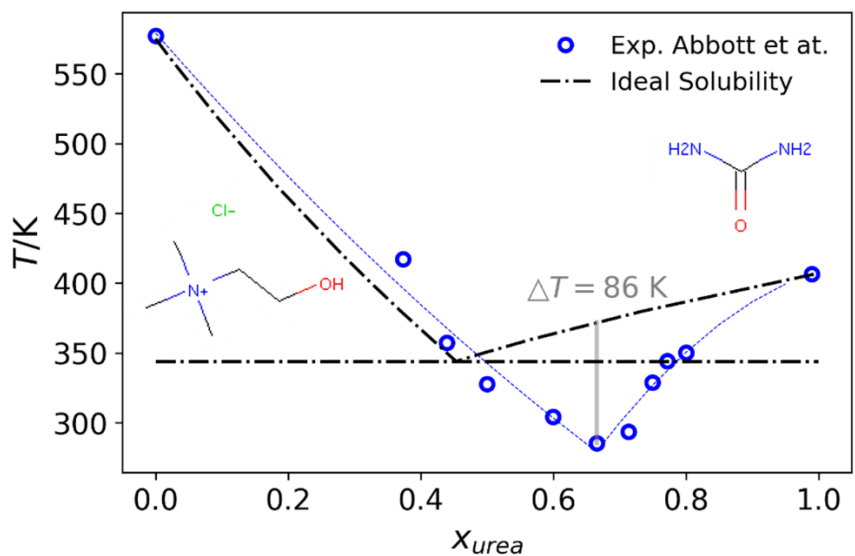

**Figure 1.** The classical ChCl + urea solid-liquid system reported by Abbott et al.[5] in 2003. The anion plays a central role as the HBA and urea as the HBD leading to attractive forces that cause a negative deviation from ideal behavior. Dashed blue line is for visual aid. Ideal solubility calculated in this work with the fusion temperature reported by Abbott et al.[5] assuming $\mathbf{\Delta C_p = 0}$ (see Table 1).

Assessing DESs as alternative solvents for different applications implies a proper description of phase equilibria. Qualitative and quantitative analyses of phase equilibria are of paramount importance for chemical process and product design[9] as well as for overcoming industrial challenges in many disciplines.[4,6] Recent thorough descriptions and reviews on the properties[1,8,10–14] and applications[1,3,4,6,15] of DESs can already be found in the literature. Therefore, the present review focuses on the thermodynamic modeling of DESs and DES - containing systems. This work covers exclusively thermodynamic modeling of DESs with excess Gibbs free energy models ($g^E$-models), equations of state (EOS) and molecular dynamics (MD) simulations.

After some general comments on the modeling of DES systems, this review continues with a section discussing the modeling of equilibrium properties such as vapor-liquid equilibrium (VLE), liquid-liquid equilibrium (LLE), solid-liquid equilibrium (SLE) at the eutectic point and modeling the solubility of a solid in a DES. Subsequently, the latest research in thermophysical and dynamic property calculations is presented. The last section summarizes further MD simulations of the liquid structure in DES systems as well as specific applications of MD calculations.

Throughout this work, the terms "deep eutectic system" and "deep eutectic mixture" are seldom used and both equally refer to a DES. In the supporting information, the reader can find a list of the reviewed literature containing the systems that were reported, the models that were applied and the DES components involved.

2. General Modeling

2.1. gE-Models and Equations of State

In this section, the generalities of the modeling of phase equilibria with $g^E$-models and EOS are described. Some described modeling approaches are taken as default and only if the modeling approach reviewed is different from the default then the details will be discussed. In most cases the modeling strategies with $g^E$-models and EOS follow textbook approaches and perhaps the only relevant modification is the handling the DES as a pseudo-component. Therefore, model-specific details will not be discussed in this section and only general modeling strategies will be mentioned.

In general, two approaches have been used to model a DES. In the first approach, the DES is not treated as its separate constituents, namely the HBD and the HBA, but rather assumed to be one indivisible pseudo-component. This has the advantage of easier handling and parametrization due to the reduction of the number of species. However, it leads to defining every different ratio of HBD to HBA as a new DES and avoids dealing with electrolyte theory. Another possibility is to model the DES as a mixture of its two constituents, having to account for the strong hydrogen bonding between these with the model. Most publications model DES as pseudo-component. This is taken as default in our review and only if the corresponding papers use the individual constituent approach, this is mentioned. The only exception is the calculation of the eutectic SLE for DES forming compounds where the individual constituent approach evidently applies. For the specific case of calculations performed with the predictive Conductor-like Screening Model for Real Solvents (COSMO-RS)[16–18], the individual constituent and the pseudo-component approach may be applied. However, it must be clarified that for the individual constituent case when a salt is present COSMO-RS may consider an electroneutral approach (cation + anion, as separate species) or a neutral ion pair approach.[19] It must be pointed out that in the reviewed literature most models apply the individual constituent approach as it is considered to be a more physical representation of the system[19].

In cubic equations of state, the needed critical properties and the acentric factor are commonly estimated when not available in literature.[20] This is because most DESs include ionic liquids (ILs) for which the critical properties commonly cannot be measured due to thermal decomposition. With other equations of state the parameters are usually adjusted to vapor pressure data of the pure DES and to liquid density data of the mixture of the DES with another solvent.

### 2.2. Vapor-liquid equilibria

For modeling vapor-liquid-equilibria in DES systems two approaches apply: the $\gamma$ - $\varphi$ concept and the $\varphi - \varphi$ concept. Most of the time the cubic equations of state are used with the $\gamma$ - $\varphi$ approach. As DESs usually have relatively small vapor pressures this approach is straightforward, since the vapor phase can be assumed to exist only for the solute (for which parameters for the EOS are usually available). In the reviewed literature the assumption that the DES is not present in the vapor phase will be taken as default and only if the systems were modeled differently the details will be discussed. This is a reasonable assumption for ionic systems, however, it could be questionable particularly when volatile HBAs or HBDs are present. Dietz et al.[21] measured the total vapor pressures in non-ionic DES systems where some of the applied DESs contained components like menthol and thymol. It was found that the vapor pressure of the DES forming components was close to zero at room temperature and in the order of 100 to 200 times smaller than that of the volatile target component in the mixture (toluene) at 100°C. The authors confirmed that the vapor pressure of the evaluated DESs were low when compared to common organic solvents. Thus, as long as a the characteristic HBA-HBD interactions of a DES are present, one may safely assume negligible DES concentrations in the vapor phase. Due to this reason most research so far has focused on the description of solubilities of gases in DES as opposed to the VLE of the DES itself. For the description of the DES in the liquid phase it is therefore a common practice to parameterize a $g^E$-model with the experimental data.

In the $\varphi - \varphi$ approach, both phases are described by an EOS. This approach usually demands a higher precision for the description of the liquid phase. For cubic equations of state, the mixing rule usually employed is the two-parameter van der Waals rule and for SAFT (Statistical Associating Fluid Theory[22])-based models the one binary interaction parameter Lorentz-Berthelot mixing rule is employed. These mixing rules are taken as the standard and if the reviewed paper uses a different approach, then this is mentioned in more detail.

### 2.3. Liquid-liquid equilibria

Liquid-liquid equilibria are commonly described by:

$$x_i^\alpha \gamma_i^\alpha = x_i^\beta \gamma_i^\beta \tag{1}$$

where the activity of species $i$ in phase $\alpha$ equals its activity in phase $\beta$. The partition of the components between the phases $x_i^\alpha/x_i^\beta$ is commonly used to assess the efficiency of an extraction process. By introducing the definition of the activity coefficient as $\gamma_i = \varphi/\varphi_i^0$ where $\varphi_i$ and $\varphi_i^0$ are the fugacity coefficients of species $i$ in the liquid mixture and the reference state, Equation (1) can also be written in the way usually used to calculate the equilibria with EOS:

$$x_i^\alpha \varphi_i^\alpha = x_i^\beta \varphi_i^\beta \quad (2)$$

LLE correlations are usually performed with the pseudo-component approach adjusting binary interaction parameters. Some predictive approaches with COSMO-RS have also been performed. In this case the details of handling the DES as a pseudo-component, an electroneutral mixture (ions + HBD) or with the ion pair approach (ion pair + HBD) will be discussed.

2.4. Solid-liquid equilibria

The SLE behavior of a eutectic mixture is commonly modeled with the following relation:

$$\ln(x_i \gamma_i) = \frac{\Delta H_m}{RT} \cdot \left(\frac{T}{T_m} - 1\right) + \frac{\Delta C p_m}{R} \cdot \left(\frac{T_m}{T} - \ln\left(\frac{T_m}{T}\right) - 1\right) \quad (3)$$

which is derived from equilibrium thermodynamics. In equation (3) $\gamma_i$ and $x_i$ are the activity coefficient and mole fraction of component $i$, respectively. The right-hand side of the equation is given in terms of the temperature of the system $T$ and the pure component melting properties of $i$: enthalpy of fusion $\Delta H_m$, change in calorific capacity $\Delta C p_m$ and melting temperature $T_m$.

Several assumptions can be made for this modeling approach. Firstly, it is assumed so far that the SLE takes place between the liquid phase and a pure ideal solid (either component $i$ or $j$) throughout the entire concentration range. This assumption holds for the modeled eutectic and deep eutectic mixtures reviewed in this work. Secondly, it is common practice to assume that the change in calorific capacity $\Delta C p_m$ can be neglected[9], leading to:

$$\ln(x_i \gamma_i) = \frac{\Delta H_m}{RT} \cdot \left(\frac{T}{T_m} - 1\right) \quad (4)$$

If the pure component properties are known, then the composition of the liquid at different temperatures below $T_m$ can be determined and the liquidus lines are obtained.

## 2.5. DESs as electrolyte systems

Most of the recent modeling approaches involving phase equilibria avoid treating IL/salt-based DESs as electrolytes. Whether it be empirical parameter fitting with EOS and $g^E$-models like the non-random two liquid (NRTL) model[23] or more predictive models (e.g. UNIFAC[24] and COSMO-RS), a specific electrolyte term is almost never included. There are of course few exceptions,[25] however these apply the unmodified Pitzer-Debye-Hückel coulombic term[26] for long-range interactions. This is unsuitable due to the decay of molecular correlations in highly concentrated electrolytes. It has even been suggested[27] that a normal NRTL outperforms the electrolyte-NRTL model[28,29] when correlating DES systems, this being attributed to the special nature of DESs. However, the shortcomings of the traditional Debye-Hückel theory offer a more plausible and descriptive explanation. This theory was developed under a framework that is valid for spherical ions at low concentrations and is very sensitive to low dielectric constant values, which would be present in many DES containing phases and pure DESs. Furthermore, considerable deviations from the theory are added on top of the aforementioned sensitivity: DESs imply large polyatomic non-spherical electrolytes in a highly concentrated regime where non-local electrostatics play a major role.[30,31] In addition, only the dielectric constant of the solvent is considered for the case of miscibility between an ionic liquid and a solvent[32]. Thus, unmodified Debye-Hückel-based terms may contribute poorly to a better description of a DES or an IL, and may even deteriorate the performance of the model that includes them, as they are being forced by "extreme" conditions unaccounted for in their derivation. Consequently, more refined electrolyte terms are needed.

Among its shortcomings, the traditional Debye-Hückel theory treats the ion cloud as bare charges instead of dressed ions (each with its own ion cloud) and does not account for the behavior of the dielectric function within its derivation. Some of these issues can be approached on theoretical grounds as shown, for instance, by dressed ion theory[33,34] and by an extended Debye-Hückel theory published by Shilov and Lyashchenko.[35–37] The influence of binary symmetric and higher order ion pairs is evidently an additional complexity to be dealt with.[37–39]

While it is still too early to provide specific directions from the chemical engineering modeling perspective, we expect these issues to be systematically addressed in the near future through the modification of transparent theories like Debye-Hückel and the primitive Mean Spherical Approximation (MSA); as has been done very recently for ILs by means of an extended Pitzer-Debye-Hückel in the work of Chang and Lin[40] that accounts for the dielectric constant of the ILs, and a modified Debye-Hückel with similar foundations to that of Shilov and Lyashchenko´s derivation[35] in the work of Bülow et al.[41] The reader is also referred to an exact theory for the decay of intermolecular correlations in electrolytes that was recently published by Kjellander.[31]

In essence, while some of the models used in the literature are flexible enough to correlate data with either the pseudo-component or the electroneutral approach, excluding coulombic interactions makes them unsuitable when it comes to a wider and consistent applicability, as will be discussed in the upcoming sections.

### 2.6. Molecular Dynamics: Force Fields for DESs

The results of MD simulations are generally determined by the quality of the applied force field, making the choice of the force field model a crucial part of MD simulations. To the best of our knowledge no such summary of force fields for DESs exists and the recent developments are discussed in detail in the following section. Since many DESs consist of an IL as HBA combined with a HBD the developments of force fields suitable for DESs are mainly based on parameters and techniques used for the simulation of ILs. In previous relevant molecular dynamics studies[42–50], the intermolecular attractions were found to be overestimated for force fields using an integer net charge of $\pm 1e$ for the IL molecules. A charge scaling of $\pm 0.8e$ for IL molecules was firstly introduced by Bhargava and Balasubramanian[51] by considering a significant charge transfer between the ions as well as polarization effects. This refinement resulted in a noticeable improvement when predicting the diffusivities for [bmim][PF$_6$]. The use of reduced net charges has become popular in the force field development and has been applied for a variety of ionic systems as it yields results that are in better agreement with experimental findings and ab initio MD simulations.[43,44,52–57] The scaling of the ionic charges in non-polarizable force fields may deliver a promising alternative to computational demanding polarizable force fields, but it needs to be adjusted to experimental data or density functional theory (DFT) calculations. As more computational power became available, polarizable force fields have grown in popularity during the past years. For more information about the state of the art of polarizable force fields for MD simulations of IL, the reader is referred to the recent review of Bedrov et al.[58].

The development of force fields suitable for a DES started with the work of Perkins et al.[59,60]. Based on modifications of the General Amber Force Field (GAFF)[61], which has previously been applied to ILs[42,43], evaluations and developments of force fields have been made mainly for ChCl-based DESs. All parameters for the bonded and intramolecular interactions are adopted from the original GAFF[61] and partial charges for the DESs have been taken from a restrained electrostatic surface potential (RESP) calculation of each molecule in an isolated state.[59,60] To better reproduce the observed charge delocalization in the IL, ionic charges of $\pm 1e$, $\pm 0.9e$ and $\pm 0.8e$ as well as different intramolecular parameters for ChCl and urea have been tested for common DESs composed of ChCl+urea[59], glycerol, ethylene glycol and malonic acid.[60] Whereas the variation of the parameters for urea and ChCl resulted in little changes, the partial charge scaling remarkably affected the overall performance of calculating the densities, molar heat capacities and volume expansivities. However, no universal net charge scaling factor could be defined, given that $\pm 0.8e$ was found to be superior in the case of ChCl+urea[59] and $\pm 0.9e$ for other DESs.[60]

Although the reduction of the partial charges has been widely accepted for ILs[43,44,52–57] and resulted in a better representation of the bulk phase properties of DESs, inconsistencies in the developed force field models were reported by Mainberger et al.[62] Using a similar approach based on GAFF[63], they extended its applicability to betanine as HBA as well as levulinic acid and 1,4-butanediol as HBD. The net charges of $\pm 0.75e$, delivering the best overall agreement with experimental thermophysical properties, significantly differ from the work of Perkins et al.[59,60] with the best overall agreement using $\pm 0.9e$. Even when the same partial charges and parameters reported by Perkins et al.[59,60] were used, their results could not be reproduced in GROMACS[64] by Mainberger et al.[62]. The main difference in the resulting charge scaling may be based on different charge assignation methods. Contrary to Perkins et al.[59,60], Mainberger et al.[62] implicitly included a charge transfer and polarization effects between the HBA and HBD by calculating the partial charges for minimal clusters regarding the eutectic composition[65,66] instead of isolated molecules. Furthermore, they suggest to adjust the entire net charge scaling by calculating the partial charges in different larger clusters in order to better reproduce the charge distribution present in the bulk phase of an IL.[47]

In addition to GAFF, Mainberger et al.[62] also investigated the Merck Molecular Force Field (MMFF).[67] In terms of MMFF no net charge scaling was necessary to reproduce thermophysical properties such as the liquid density and heat capacities with quantitative agreement compared to experimental findings. However, similar challenges related to dynamic properties of DESs previously reported by Perkins et al.[59,60], are also observed for MMFF and GAFF[62] (further details in section 5).

Besides the development of force fields capturing the properties of different DESs, the refinement of force field parameters for one specific DES application by Ferreira et al.[68] and García et al.[66] have led to valuable insights into modeling of complex DES systems using MD simulations. Whereby, García et al.[66] studied the influence of the charge distribution between choline, chloride and levulinic acid while taking all other parameters from a previous study.[69] They calculated the partial charges of every species in minimal clusters according to the eutectic mixture as well as isolated molecules with DFT methods. A large dependency of the used charge assignation scheme on the resulting charge distribution was found. A charge spreading from the chloride ion towards levulinic acid was found in every charge assignation method in accordance with ab initio MD simulations for other organic solvents[65]. The authors recommend the charge assignment based on electrostatic potentials (e.g., ChelpG[70] or Merz-Kollman[71]) combined with minimal cluster calculations for the development of force fields for DESs, as they give a reasonable depiction of the nanostructure of the investigated solvent. In relation to all charge assignation methods investigated by García et al.[66], the optimal charge distribution represents a significant charge transfer between the ions and the HBD. Ferreira et al.[68] on the other side compared a variety of combinations of parameters and partial charges from the literature[59,60,72–77] in order to find an improved force field for ChCl+ethylene glycol. In accordance with previous findings the densities were accurately predicted by the original force fields. However, calculating dynamic properties such as the diffusion coefficient required refinements to account for the effect of electronic polarization. In contrast to prior studies, which determined the charge distribution in isolated molecules or small clusters, the rescaling of the partial charges has been compared to bulk phase DFT calculations resulting in best agreement with the potential obtained with $\pm 0.8e$. Using these scaled charges, the performance of all tested force fields significantly improves, in particular for self-diffusion coefficients. Ferreira et al.[68] found the overall best agreement with experimental findings by combining force field parameters developed by Sambasivarao and Doherty[77] for the choline ion, the original OPLS-AA parameters for the chloride ion and parameters of Szefczyk and Cordeiro[74] for ethylene glycol with the reduced ionic charges of $\pm 0.8e$.

The previously discussed force fields were optimized towards representing a particular DES. Lacking of a systematical refinement of force fields in order of covering a variety of DESs, Doherty and Acevedo[78] developed new parameters for many ChCl-based DESs on the basis of their OPLS-AA force field for IL.[79] The authors showed in prior studies[79,80], that ionic charges of $\pm 0.8e$ in the OPLS-AA force fields for IL were found to provide the best representation of both bulk and local intermolecular properties. This is in good agreement with the DFT calculations of Ferreira et al.[68]. As MD simulations have proven to be sensitive towards the calculation of partial charges, Doherty and Acevedo fitted the charge distribution in the DES molecules to experimental derived radial distribution functions. The non-bonded parameters are subsequently adjusted to reproduce bulk phase properties and ab initio RDFs resulting in the OPLS-DES[78] force field. This force field represents a remarkable quantitative agreement with several thermophysical properties (errors of 1-5%) and is applicable to a variety of HBDs.

In summary, existing force fields have been refined towards handling DESs in MD simulations. The advantage of refined force fields originally developed for bio-molecular systems (e.g., GAFF[59,60,62], OPLS-AA[68,78]), is that the interactions of a DES with other molecules such as proteins can be studied. The observations of different research groups prove that the overestimation of the interionic interactions in DESs is not unique to one specific force field but a typical problem in simulating ILs and DESs with non-polarizable force fields. Therefore, scaling the partial charges of the ionic components was found to be a general treatment necessary to reflect the average polarization state of a DES in a non-polarizable force field. Although no universal charge scaling parameter could be determined within the force field development, the best performance was achieved using ionic charges in the range of $\pm 0.75e$ to $\pm 0.9e$, which is in good agreement with some previous force field developments for ILs.[51,57] While the prediction of dynamic properties was proven to be challenging (see section 5), another drawback of charge scaling in non-polarizable force fields may be the inadequate calculation of phase equilibria of the IL components. Son et al.[81] have shown that predictions of the phase behavior in ILs may not be reasonable, if scaled charges are used. As many DESs consist of an IL, this drawback may also apply for the prediction of phase equilibria of DES using MD simulations. This is a question for future research. Besides some drawbacks (see sections 5 and 6), MD simulations using non-polarizable force fields aid in gaining deep insights into the intermolecular interactions and molecular behavior of a DES. Regarding the developments of polarizable force fields for ILs, much progress has been made in the past years.[58] In line with the rising availability of computer power, the application of these polarizable force fields to study DES systems may become a topic of particular interest in the near future.

3. Activity and Phase Equilibria

This section reviews the description of different equilibrium properties of DES systems. The literature is discussed in the following order: $g^E$-models and EOS, MD simulations. Additionally, an overview of all modelled systems and corresponding properties is provided in Table S1 the supporting information.

3.1. Activity Coefficients

Although activity coefficients are mostly used to calculate phase equilibria, for example VLE or LLE, some studies report them individually. This has been done with the purpose of screening different solvents by the use of infinite dilution activity coefficients. In addition, water activity can be related to protein properties. Wedberg et al.[82] found a water activity influence on the hydration layer around enzymes and therefore water may play an important role in ensuring the enzymatic stability in non-aqueous media.

The PR-EOS has been used to calculate the water activity[83] in aqueous mixtures of DES based on ChCl + glucose. Verevkin et al.[84] applied PC-SAFT to predict infinite dilution activity coefficients and get an estimate of the relative solubility of different hydrocarbons and aromatic hydrocarbons. PC-SAFT in most of the cases was able to predict successfully the order of magnitude and also correctly predict the temperature trend in all cases. Baz et al.[85] compared the activity coefficients for ChCl+glycerol mixtures with water obtained from PC-SAFT by modeling the DES as one pseudo-component or as two components. The results were in good agreement with experimental activity coefficients when modeling the DES with the individual constituent approach.

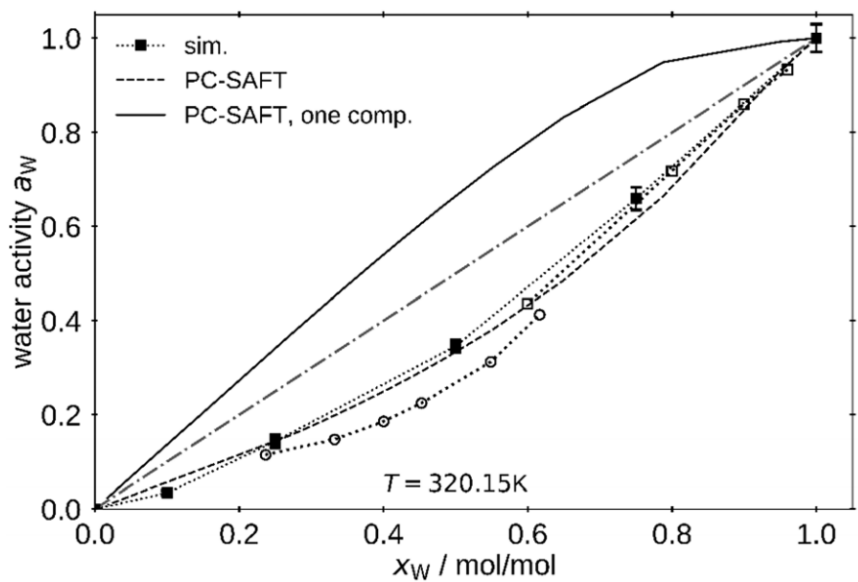

**Figure 2.** Activity of water in mixtures with ChCl+glycerol calculated by MD simulations (closed symbols with dotted line) and from PC-SAFT (solid and dashed line). Dash dotted line represents

the ideal mixture. Experimental values are illustrated as open circles[86] and open squares[87]. (Republished with permission of The Royal Society of Chemistry, from Baz et al.[85]; permission conveyed through Copyright Clearance Center, Inc. Elsevier)

Besides the calculation of activity coefficients using an EOS or a $g^E$-model, they can also be calculated with MD simulations from solvation free energies by the use of thermodynamic integration.[82,88,89] Although this methodology is more computationally demanding in comparison to other methods, it might be advantageous if the activity of water can be determined in the presence of an enzyme. Studying the activity coefficients of water in a ChCl+glycerol - water mixture, Baz et al.[85] observed negative deviation from Raoult's law (Figure 2), which denotes a stronger interaction of water with the DES mixture compared to pure water. Their simulated results are in fair agreement with different experimental measurements. However, the robustness of the calculation of the activity coefficients from MD simulations has been questioned by Baz et al.[85], as the activity coefficients show a large sensitivity towards the free-energy difference determined by MD simulations.

3.2. VLE

VLE calculations generally assume that the DES components do not have a significant concentration in the gas phase. This is a safe assumption as long as the characteristic HBA-HBD interactions are present in the liquid phase.[21]

**$\gamma$ - $\varphi$ concept**

One of the first to employ this modelling approach without the use of an equation of state i.e. assuming the fugacity coefficient equal to unity was Jiang et al.[90]. They evaluated the use ChCl-based DESs as azeotrope breakers for the isopropanol + water VLE. A COSMO-RS implementation was applied for solvent screening in order to select the best possible DES based on infinite dilution activity coefficients. Experimental data was correlated with the NRTL model. This group followed the same modelling approach for other systems composed of DES + water + allyl alcohol.[91] In both cases the DESs were considered as a pseudo-component (pseudo-ternary approach).[90,91]

In contrast, Zhang et al.[92] applied the individual constituent approach . They evaluated the water + 2-propanol system at different DES (ChCl+glycerol) concentrations and fitted an NRTL model to the experimental data. These authors also assumed fugacity coefficients and pointing factor equal to unity.

A different parameterization approach is presented by Ma et al.[93,94] by fitting an NRTL model to enthalpies of mixing for different DES aqueous systems and then applying these NRTL parameters to estimate the vapor pressure of water in the DES at different temperatures assuming fugacity coefficients equal to unity. The calculated vapor pressures were in agreement with values from the literature but presented deviations with rising temperature. Furthermore, the authors comment on the deviations for the ChCl + malonic acid DES (1:1) for which the vapor pressures of water deviate considerably as the DES concentration in the mixture rises.[93]

One of the first attempts to model DES (ChCl+Urea) systems using the $\gamma$ - $\varphi$ concept with the use of an equation of state was presented by Li et al.[95] who modeled the solubility of $CO_2$ in the DES liquid phase. Although the solubility of $CO_2$ in the liquid phase reaches a mole fraction of 0.3 for the 1:2 DES, the activity coefficient is assumed to be unity (i.e., equal to infinite dilution). By means of the Peng-Robinson (PR)-EOS to calculate the fugacities of the solute, they were able to calculate Henry constants for different temperatures.

The system $CO_2$ + DES (mainly ChCl+Urea) has received much attention in the literature. Ji et al.[96] analyzed this system, but focused on the influence of the water concentration on the solubility of $CO_2$ in the DES. They used a combination of the RK-EOS for describing $CO_2$ in the vapor phase and NRTL to describe the DES in the liquid phase. Xie et al.[97,98] applied the same strategy to correlate the solubility of several gases including $H_2$, $N_2$, CO, $CH_4$ and $CO_2$ in ChCl+Urea mixtures. Ma et al.[99] simulated a complete process to separate $CO_2$ from $CH_4$. Later the same group also compared the technical and economic aspects of running such a process[100] by calculating some of the proposed solvents for $CO_2$ removal with the RK+NRTL and with PC-SAFT.

Chen et al.[101], Ghaedi et al.[102] and Zubeir et al.[103] applied the PR-EOS to determine the fugacity of $CO_2$ for calculating Henry coefficients in DESs. Duan et al.[104] employed the two-terms virial equation to model the fugacity of $NH_3$ to calculate the Henry constants in DES based on ChCl.

Mirza et al.[105] used the modified alpha function for species with acentric factors larger than 0.49 in conjunction with the PR-EOS to calculate the Henry coefficients of $CO_2$ in DESs composed of ChCl+Urea or ethylene glycol and a ternary DES composed of ChCl, ethylene glycol and malic acid. For all three systems binary interaction parameters were adjusted to accurately represent the measured VLE data. Lu et al.[106] employ the two-term virial EOS and were able to represent the Henry coefficients as well. Gjineci et al.[107] used a combination of a two-term virial EOS and a $g^E$-model to correlate the DES influence on the breaking of the azeotrope ethanol/water. For the virial EOS they applied the Tsonopoulo[108] correlation to calculate the second virial coefficient. For the tested $g^E$-models (UNIQUAC and NRTL) a good correlation of the experimental data was found. For the same system, Peng et al.[27] applied the same modelling approach adjusting not only binary but also ternary NRTL parameters. Sharma et al.[109,110] employed the same modeling approach to correlate the DES effect on the VLE of the system acetonitrile/water.

Leron et al.[111] combined an EOS for $CO_2$ developed by Huang et al.[112] with a simplified Pitzer model to correlate the Henry constants. Haghbakhsh et al.[113] integrated the cubic-plus association (CPA) EOS for the vapor phase with NRTL and UNIQUAC for the liquid phase to calculate temperature dependent solubilities of $SO_2$ at atmospheric pressure. The CPA parameters were adjusted to densities and vapor pressures while the $g^E$-model parameters were adjusted to the experimental VLE data. Regardless of the $g^E$-model employed the fitted models deliver an almost equal agreement with experimental data. Haider et al.[114] compare the ability of the PR EOS with the model NRTL to fit the Henry constant and found that both models have a comparable good performance. Later on, the same group also modeled a $CO_2$/$CH_4$ capturing process[115] using the PR-EOS with quadratic temperature dependent binary interaction parameters to correlate the solubilities over a wide range of temperatures and pressures in two ChCl based DESs.

In contrast to all the previous approaches, Zhong et al.[116] successfully modeled the solubility of $NH_3$ in ternary DES as a pseudo-reaction adjusting one reaction constant to the experimentally determined values.

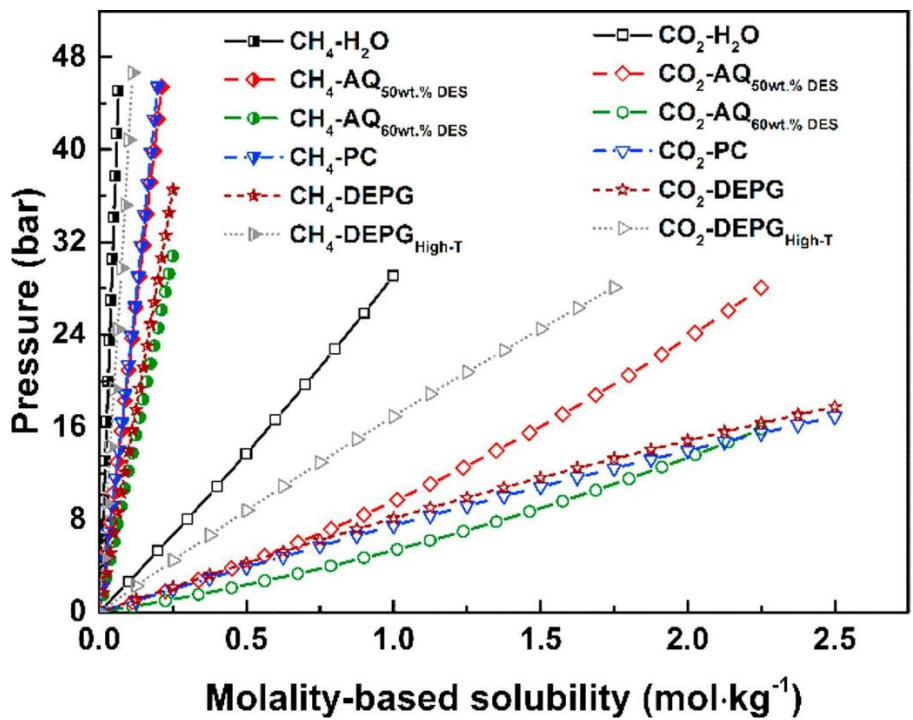

**Figure 3.** Example of solubilities of $CO_2$ and $CH_4$ at 293.15K in several solvents. (reprinted from Ma et al.[100] with permission from Elsevier)

It seems that if only the solubility of gases in DESs are the target property, the assumptions of Henry's law are applicable for most systems up to relatively high pressures. Also the assumption of the gas phase consisting only of the solute gas is close to reality; thus, the fugacity coefficient of the solute is the relevant measure to describe the non-ideality of this phase. Spycher and Reed[117] show that for the majority of DES systems studied so far, the fugacity coefficient of the solute gas up to 50 bars and even higher is close to unity. If the activity coefficient of the solute in the liquid phase is also assumed to be unity in most cases the solubilities seem to be directly proportional to the pressure (see Figure 3). Following this line of thought would mean, that the Henry coefficient is simply equal to the slope of the graph, as has already been applied successfully by Wu et al.[118] and Liu et al.[119] It does not make practical sense to increase the complexity of a model any further. At least for the systems studied so far, it seems to be a better modeling strategy to describe the liquid phase with a $g^E$-model assuming fugacity coefficients equal to unity when the solubility ceases to be directly proportional to the pressure.

**$\varphi$ - $\varphi$ concept**

To model DES systems with the $\varphi$ - $\varphi$ approach (for instance using cubic equations of state to describe both phases) it is usually necessary to know the critical properties and the acentric factor. Unfortunately in many cases the DESs thermally decompose and it is not possible to measure these quantities[120]; hence, experimental data is not available, especially at different HBD to HBA ratios. Therefore, property estimation methods are needed.

Mirza et al.[20] describe the most common methodology to estimate the needed properties of a DES: it consists of the application of the modified Lyderson-Joback-Reid group contribution method[121] to calculate the critical properties of the HBD and the HBA plus a correlation to calculate the acentric factor[122] in combination with the Lee-Kesler mixing rules as recommended by Knapp[123] for the DES. The mixing rules do not require external input other than a binary interaction parameter $k'_{ij}$ to calculate the critical temperature of the DES which in most cases is set to unity. Mirza et al.[20] also calculate liquid densities at 40 °C with an empirical corresponding EOS[124] to check the validity of the predicted critical properties based on a comparison with experimental densities. While this validation is important and there are not many other tests that can be done in this case, the applied EOS[124] was developed for pure components and tested for non-hydrogen bonding mixtures of hydrocarbons. Furthermore, for HBD with amides and HBA based on methyltriphenylphosphonium bromide the deviations range from -9% to +17% of the measured experimental densities respectively. Nevertheless, overall the calculated values agree with the experimental values within an average relative deviation of about 5%, showing that the calculated critical properties have reasonable values.

Ali et al.[125] applied the PR-EOS with the one parameter van der Waals mixing rule introducing a binary interaction parameter for the attractive parameter to model the solubility of $CO_2$ in a variety of DESs composed of different HBDs and HBAs. The qualitative trend of the solubilities was predicted correctly using no binary interaction parameters ($k_{ij}$) for most systems, which gives credit to the aforementioned method to estimate the DES properties. Interestingly, when adjusting $k_{ij}$ they showed that larger solubility results in smaller interaction parameters. In a follow-up work[126] they used the parameters from their previous publication to model a $CO_2$ capturing process. Zubeir et al.[127] also employ the PR-EOS with the two parameter van der Waals mixing rules to model the solubility of $CO_2$ in several low transition temperature mixtures (LTTMs). LTTMs behave similarly to DESs, but instead of having a first order phase transition, they show a second order phase transition. They also employ the method described earlier to calculate the critical properties of the DES. However, to calculate the critical temperature of the DES the parameter $k'_{ij}$ was taken as a function of the critical volume of the DES-building components:

$$k'_{ij} = \left[2\left(V_{c_i}V_{c_j}\right)^{1/6} / (V_{c_i}^{1/3} + V_{c_j}^{1/3})\right]^3.$$

Haghbakhsh et al.[128] investigated the performance of the CPA EOS to correlate the $CO_2$ solubility in 15 DESs trying different association schemes for $CO_2$. For each DES five parameters were adjusted to experimental densities. Finally, the authors recommend to model $CO_2$ as non-associating.

Zubeir et al.[129] were the first to model DES systems with PC-SAFT. They compare the two ways to model the DES: as a pseudo-component or as two individual constituents. In the first case, they adjusted the parameters of the DES to liquid densities leaving the association parameters fixed and in the second case, these were adjusted to liquid densities of mixtures of the DES components with water. Although the pseudo-component approach gives overall slightly better results by adjusting binary interaction parameters for different ratios of HBD to HBA, the individual component approach allows quantitative prediction requiring only one binary interaction parameter for each species. To further understand DES systems, Baz et al.[85] compare thermophysical properties for one DES (ChCl+Glycerol) calculated by PC-SAFT and MD. They conclude that the pseudo-component approach represents the densities of the DESs better, while the individual component approach calculates the water activity in mixtures of water + DES more accurately. This is not very surprising given that in the first approach the parameters are adjusted by correlating the experimental density using binary interaction parameters, while in the second approach no binary interaction parameter between the HBD and HBA is used. Furthermore, Dietz et al.[21] showed, that by introducing this binary interaction parameter between the individual DES constituents, vapor pressures for different DES could be correlated using single constituent parameters. Another SAFT -based EOS used to model DES systems is the soft-SAFT. In the first publication[130] the authors evaluated which of the two modelling approaches works best. They concluded, that although the pseudo-component approach delivers slightly better results (similar to PC-SAFT), modeling the DES as two components is more universally applicable. In their next publication[131] the authors performed a consistent parameterization of transferable molecular models for eutectic ammonium salt- based DESs to correlate the solubilities of $CO_2$ and $SO_2$ with temperature independent binary interaction parameters.

Although a variety of results point to preferring the individual constituent approach, in the literature the pseudo-component approach is commonly used. PC-SAFT has also been employed with the same parameterization approach as the work by Zubeir et al.[129] to predict the VLE for mixtures of hydrophobic DESs[132] as pseudo-components and $CO_2$ without the use of binary interaction parameters showing again the strong theoretical foundation of the model. Animasahun et al.[133] make a comparison of the Soave-Redlich-Kwong (SRK), the PR and the PC-SAFT EOS for correlating solubilities in 20 different DESs modelled as pseudo-components. The parameters for the cubic equations of state were only adjusted to the VLE data while the parameters of PC-SAFT were adjusted to experimental densities and to the VLE data. They found, that all three equations of state perform almost equally.

VLE calculations with MD are limited. In addition to a few studies testing the suitability of DESs for $CO_2$ capture[134] or $N_2$ and methane absorption[135–137], Ullah et al. [69] have investigated (ChCl + levulinic acid) for $CO_2$ absorption whereby the hydrogen bond structure of the DES was not disturbed by the enclosure of $CO_2$. Furthermore, a layer of levulinic acid was found at the interface between the DES and $CO_2$ resulting in large residence times of $CO_2$ at the interface as well as in a reduced mass transfer into the bulk fluid.

### 3.3. LLE

To the extent of our knowledge, only $g^E$-models and PC-SAFT have been applied to describe LLE systems containing eutectic solvents. The most commonly analyzed systems involve petroleum industry oriented applications like the extraction of aromatics, alcohols, thiophene or benzo-thiophene from an aliphatic phase. In these works, it seems almost an established practice to report either a fitted NRTL model, COSMO-RS based predictions, or a comparison of both.[4,138–153] Parameters for UNIQUAC have also been reported for these types of systems.[145,154] Applying COSMO-RS as a screening tool to select DESs based on the performance index (*PI*) for liquid extraction purposes is also a widespread practice and is usually well described in the literature.[138,152,155] For a critical review on the extraction of aromatics from an aliphatic phase using DESs the reader is referred to the work of Hadj-Kali et al.[4]

The calculation of pseudo-ternary LLE systems with PC-SAFT (DES / alkane / tiophene) was reported by Warrag et al.[156]. To achieve this, first the DESs were modelled with the pseudo-ternary approach adjusting their parameters to temperature dependent densities. Subsequently, the binary interaction parameters were adjusted to binary LLE data for the DES-thiophene and DES-alkane systems. PC-SAFT was able to calculate trends for the distribution coefficients in accordance with the experimental data. Thus, the standard procedure of adjusting binary interaction parameters to reproduce ternary systems can be extended to DESs using the pseudo-component approach.

For NRTL and UNIQUAC the DES is commonly modeled with the pseudo-ternary approach[4,142,143,145,152–154] whereas for the COSMO-RS methodologies the electroneutral approach (quaternary) is preferred and a recalculation of the mole fractions must be performed for the DES to present the results as pseudo-ternary.[140,147,148] In addition, modeling LLE systems with the pseudo-component approach in COSMO-RS is also reported in the literature. Gouveia et al.[144] evaluated three different DESs as azeotrope breakers for the toluene / heptane system and their results (handling the system as pseudo-ternary) are correlated with NRTL and modeled with COSMO-RS assuming the ion pair + HBD to be a single neutral component. Furthermore, Bezold et al.[148] present a thorough analysis of different COSMO-RS based modeling strategies for solute partitioning where it was found that the electroneutral and ion pair approaches give qualitatively comparable results. This gives rise to the question of why there is a good and simultaneous agreement between NRTL (pseudo-ternary), COSMO-RS (electroneutral approach) and COSMO-RS (neutral ion pair or neutral DES complex approach) with the experimental data.

The main reason for this is that the aliphatic rich phase in the analyzed aromatic / aliphatic and alcohol / aliphatic systems usually contains little to no DES components. As has been pointed out by Bezold et al.[148], assuming a constant HBD:HBA ratio in the DES-rich phase is a reasonable assumption when negligible concentrations of the HBD and HBA are found in the aliphatic phase.

The previous observation is in agreement with results from the literature and can be supported with insights obtained from experimental data. Wazeer et al[138] studied the LLE formed in benzene / alcohol systems after the addition of ChCl-based DESs with the objective of evaluating the performance of the alcohol extraction process. In their experimental work they noted that when the alcohol exceeds ~40% wt in the feed, the biphasic system tended to become a one phase system. The authors attributed this behavior to a weakening of the HBA-HBD interaction of the DES due to the increasing alcohol concentration. Empirical evidence from 2D NMR (NOESY and HOESY) analyses were performed to support this observation and the results suggested that above a 40 wt % of alcohol the hydrogen bonding of the DES weakens and eventually disappears as the alcohol concentration increases. The same group had similar findings when studying the potential of DESs in enhanced oil recovery (EOS).[157]

A recent example is provided by Bezold and Minceva[158] (shown in Figure 4) for the biphasic system formed by a n-heptane / methanol / DES where the DES is L-menthol + levulinic acid. The authors stress the fact that even though a liquid DES was added to the mixture the individual components of the DES actually distribute between the liquid phases. Thus, arbitrarily assuming a constant HBD:HBA ratio in a phase may be unphysical in many cases and the system must be characterized as quaternary.

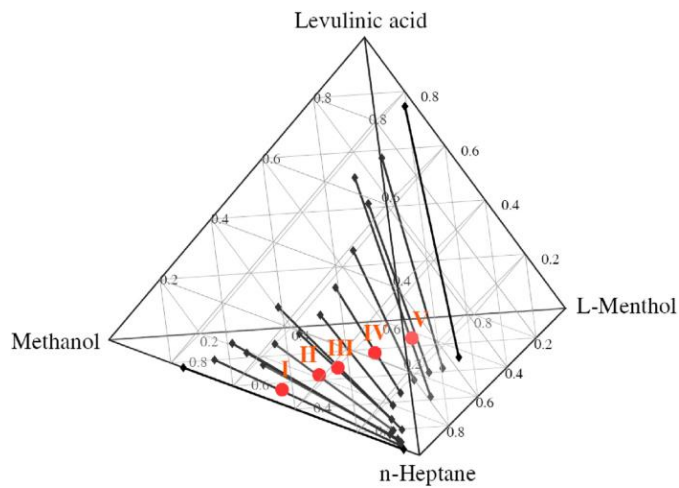

**Figure 4.** Liquid-liquid phase diagram of n-heptane/methanol/L-menthol/levulinic acid. Quaternary plot reprinted from Bezold and Minceva[158] with permission from Elsevier.

Regarding other types of systems (not involving an aliphatic phase), experimental data and NRTL parameters have been reported in the literature for alcohol-ester systems where a DES is assessed as an entrainer[159,160] for alcohol extraction purposes. Rodríguez et al.[161] report the experimental data for pseudo-ternary systems where glycerol-based DESs were applied as extractants for methyl-ethyl-ketone (MEK) / ethanol mixtures and the data was correlated with the NRTL model. In their results it can be observed that the DES has a low concentration in the MEK rich phase. The NRTL model fits the data and the authors recovered the DES and analyzed its integrity via H-NMR.[161]

A different approach for COSMO-RS based modeling is presented by Verma and Banerjee[162]. In their work the partition of alcohols (ethanol, propanol, butanol) between an aqueous and a non-ionic DES phase was measured, correlated with the NRTL model and compared with COSMO-RS predictions. In this case the DES (DL-menthol + lauric acid) was not modeled by generating the *cosmo file* of a paired structure. Instead, an averaging of the sigma profiles based on the molar ratio (2:1) of the DES components was performed. COSMO-RS satisfactorily reproduced the qualitative trends for all systems.

Samarov et al.[159] studied the use of ChCl-based DESs as entrainers for alcohol-ester systems also handling the systems as pseudo-ternary and fitting the results to an NRTL model. The authors comment on the quaternary nature of the system and, even though the pseudo-ternary treatment simplifies the modeling approaches, they consider alcohol concentrations below 50 wt % and the stability of the DES structure in both phases was confirmed. Thus, validating the pseudo-ternary approach.

Finally, Liu et al. evaluate ChCl-based DESs for the separation of methanol[163] and ethanol[164] from dimethylcarbonate and correlate their experimental results with an NRTL model. The pseudo-ternary approach is validated by confirming the integrity of the DES in both phases via H-NMR.

Using the pseudo-ternary approach Dietz et al.[165] also correlated liquid-liquid equilibria of DES+water and solubilities of HMF (5-hydroxymethylfurfural) in DESs with the PC-SAFT EOS. Furthermore, they were also able to perform qualitative predictions[166] of the infinite dilution partition coefficient of furfural and HMF in mixtures of DESs and water. The correlations were improved further by adjusting binary interaction parameters.

As a general conclusion, the pseudo-ternary approach can be assumed to be valid for thermodynamic modeling of LLE systems when the concentration of the DES in one of the phases is negligible. In the opposite case, the pseudo-ternary approach still remains a reasonable assumption if the integrity of the DES is empirically confirmed in both phases. Moreover, the case of a DES with ionic components implies that the partition of the ions in the phases must eventually be accounted for. A characterization of ion-solvent interactions in the mixture is required if the phenomena that take place are to be properly modeled. This in turn foreshadows an additional obstacle: experimental LLE with DESs are commonly reported as pseudo-ternary systems. Conversely, quaternary LLE experimental data is very scarce in the DES literature.

### 3.4. SLE: eutectic points

One of the salts that is very commonly applied to produce DESs is ChCl. However, it is not stable at high temperatures and its fusion properties cannot be measured.[1] One of the most widely applied approximations of these is reported by Fernandez et al.[167] The quasi-ideal behavior of selected ChCl-based eutectic systems was assessed via equation (4) applying COSMO-RS to calculate the activity coefficients. These were compared with experimental data and an indirect assessment of the fusion properties of ChCl was performed.[167] The results obtained by the authors for ChCl and the fusion properties of common components found in the reviewed DES literature for SLE calculations are presented in Table 1. The reader may find additional fusion data for quaternary ammonium chlorides, bromides, iodides, nitrates and perchlorates, among others in the work of Coker et al.[168] It must be kept in mind that some of these might be biased by the decomposition temperature of the salts. Marcus has also reported fusion data for unconventional DESs formed by salt hydrates and water[169]. Finally, fusion data of miscellaneous components can also be found online in the Chemistry WebBook and the Ionic Liquids Database (IL Thermo v2.0) of the National Institute of Standards and Technology (NIST).

Reliable values for the fusion properties of DES -forming components are generally scarce due to the thermal instability of many of the involved organic compounds. Consequently, SLE modeling of the eutectic point with gE-models and equations of state is relatively scarce[2] and methodologies like the work of Fernandez et al.[167] as well as experimental efforts should be encouraged.

To the extent of our knowledge, only the PC-SAFT equation of state and the NRTL, Redlich-Kister (RK), UNIFAC(Dortmund) and COSMO-RS $g^E$-models have been applied to describe the SLE behavior of deep eutectic mixtures.

Pontes et al.[170] present solid-liquid phase diagrams for the eutectic systems of three quaternary ammonium salts and their combination with different fatty acids. The experimental data is successfully correlated with the PC-SAFT equation of state and all the systems exhibit a clear negative deviation from ideal solubility. This negative deviation was found to be more pronounced as the alkyl chains of the quaternary ammonium cation increased in length.

Martins et al.[7] also present results with PC-SAFT for eutectic mixtures formed by mixing terpenes and monocarboxylic acids. The authors found a good agreement with the EOS, nevertheless most systems exhibited a quasi-ideal behavior; whereas in another publication[171] eutectic mixtures formed between ChCl and carboxylic acids showed considerable negative deviations from ideal behavior. In the second case, PC-SAFT is able to describe these non-idealities.

**Table 1.** Fusion properties of eutectic forming compounds commonly found in the DES literature.

| Component | $\Delta H_m$/kJ mol$^{-1}$ | $T_m$/K | Taken from: |
|---|---|---|---|
| 1-hexadecanol | 60.96 | 324.40 | Crespo et al.[172] |
| 1-octadecanol | 65.35 | 332.90 | Crespo et al.[172] |
| 1-tetradecanol | 45.81 | 311.70 | Crespo et al.[172] |
| Benzoic Acid | 18.01 | 395.52 | Wolbert et al.[173] |
| Benzyldimethyl(2-hydroxyethyl)-ammonium chloride | 8.73[a] | 351.42[a] | Fernandez et al.[167] |
| Capric Acid | 27.50 | 304.75 | Pontes et al.[170] |
| Caprylic Acid | 19.80 | 288.20 | Martins et al.[7] |
| Choline Acetate | 8.882[a] | 362.62 | Fernandez et al.[167] |
| Choline bis(trifluoromethylsulfonyl) imide [Ch][NTf$_2$] | 1.227 | 305.65 | Fernandez et al.[167] |
| Choline Butanoate | 8.794 | 315.98 | Fernandez et al.[167] |
| Choline Chloride (ChCl) | 4.30[a] | 597.0[a] | Fernandez et al.[167] |
| | | 575.15[b] | Abbott et al.[5] |
| Choline Propanoate | 2.239[a] | 282.57 | Fernandez et al.[167] |
| Cinnamic acid | 22.21 | 406.10 | Wolbert et al.[173] |
| Citric Acid | 40.32 | 427.80 | Crespo et al.[171] |
| D-Fructose | 33.0 | 386.75 | Silva et al.[2] |
| D-Glucose | 32.0 | 420.0 | Silva et al.[2] |
| DL-lactic acid | 11.34 | 289.80 | Crespo et al.[171] |
| DL-menthol (α polymorph) | 14.20 | 306.95 | Corvis et al.[174] |
| D-Manose | 24.69 | 407.0 | Silva et al.[2] |
| D-Xylose | 31.65 | 423.0 | Silva et al.[2] |
| Glutaric Acid | 20.70 | 370.6 | Crespo et al.[171] |
| Glycolic Acid | 19.30 | 350.8 | Crespo et al.[171] |
| L-arabinose | 35.78 | 435.0 | Silva et al.[2] |
| Lauric Acid | 37.83 | 317.48 | Pontes et al.[170] |
| | 34.70 | 317.90 | Wolbert et al.[173] |
| Lidocaine | 16.40 | 340.65 | Wolbert et al.[173] |
| L-menthol (α polymorph) | 14.10 | 316.05 | Corvis et al.[174] |
| L-menthol | 12.89 | 315.68 | Martins et al.[7] |
| Malic Acid | 25.30 | 403.15 | Martins et al.[8] |
| Mannitol | 54.69 | 437.3 | Martins et al.[8] |
| Meso-erythritol | 38.9 | 391.20 | Martins et al.[8] |
| Mystiric Acid | 41.29 | 327.03 | Pontes et al.[170] |

|  |  | 45.75 |  | Martins et al.[8] |
| --- | --- | --- | --- | --- |
| Oxalic acid |  | 18.58 | 462.40 | Crespo et al.[171] |
| Palmitic Acid |  | 51.02 | 336.84 | Martins et al.[7] |
| Paracetamol |  | 18.59 | 349.15 | Wolbert et al.[173] |
| Para-toluic acid |  | 22.72 | 452.80 | Wolbert et al.[173] |
| RS-Ibuprofen |  | 25.50 | 350.24 | Wolbert et al.[173] |
|  |  |  | 348.75 | Corvis et al.[175] |
| S-Ibuprofen |  | 18.60 | 324.75 | Corvis et al.[175] |
| Stearic Acid |  | 61.36 | 343.67 | Martins et al.[7] |
| Succinic Acid |  | 32.95 | 460.70 | Crespo et al.[171] |
|  |  | 34.0 | 455.20 | Martins et al.[8] |
| Sucrose |  | 32.0 | 462.05 | Silva et al.[2] |
| Tartaric Acid |  | 36.31 | 443.30 | Crespo et al.[171] |
| Tetrabutylammonium Bromide |  | 15.48[b] | 395.15[b] | Coker et al.[168] |
| Tetrabutylammonium Chloride |  | 19.43[a] | 342.82[a] | Fernandez et al.[167] |
|  |  | 20.50[b] | 314.15[b] | Coker et al.[168] |
| Tetraethylammonium Chloride |  | 51.24 | 526.78 | Pontes et al.[170] |
| Tetramethylammonium Chloride |  | 20.49 | 612.87 | Pontes et al.[170] |
| Tetrapropylammonium Chloride |  | 66.58 | 503.07 | Pontes et al.[170] |
| Thymol |  | 19.65 | 323.50 | Martins et al.[7] |
| Urea |  | 14.6 | 407.20 | Martins et al.[8] |
| Vainillin |  | 22.40 | 355.40 | Wolbert et al.[173] |

[a] Estimated Values. [b] Values possibly biased by the decomposition temperature.

The aforementioned group also evaluated the SLE behavior of ChCl-based eutectic systems formed with fatty acids and fatty alcohols.[172] The results presented in their work were also correlated with the NRTL model (with no temperature dependency). Though both models describe the eutectic systems satisfactorily, the PC-SAFT equation of state outperformed the NRTL model in describing the systems that present negative deviations from ideal behavior. The authors justify this by the fact that the NRTL model has no explicit term for the strong hydrogen bonding interactions that are characteristic of a deep eutectic mixture.[172] Such an observation is in agreement with the modeling results from Silva et al.[2], where the NRTL and temperature dependent NRTL models were applied to describe ChCl+sugar-based deep eutectic mixtures. However, a modified-RK expansion was found to be more suitable to represent the negative deviations from ideal solubility (on average the objective function attained values 55% below those of NRTL). This modified-RK expansion was then used for the reference values to train a modified COSMO-RS by optimizing the hydrogen bonding interaction and temperature dependent terms as well as the misfit prefactor. The authors state in general terms that COSMO-RS underestimates the non-ideality of the sugars and overestimates the non-ideality of ChCl.[2] The methodology is detailed by the authors in the supplementary materials of their publication.

Regarding the modified-RK and NRTL correlations as well as the COSMO-RS based predictions, modeling the salt as a neutral ion pair is the preferred method for eutectic point calculations in DES systems.[2,167,172] In spite of the fact that these are in many cases highly concentrated electrolyte systems, it seems so far more convenient to treat them as a mixture of neutral associated species for SLE calculation purposes. The implications of this are beyond the scope of the present review, but it is our belief that questions regarding the non-local electrostatics and structure of highly concentrated electrolytes like DESs and ILs must be addressed.[13,30,176] Furthermore, in order to use equation (3) a non-conventional procedure to change the reference state when a liquid mixture is treated as an electrolyte solution is required.[177,178]

Even though many eutectic systems reported in the DES literature contain ionic species, there is also a growing interest in non-ionic DESs[1]. One example is the use of therapeutic deep eutectic systems (THEDES). THEDES are obtained by mixing an active pharmaceutical ingredient (API) with an excipient. The objective is enhancing the bioavailability of the API by stabilizing its amorphous form in a thermodynamically stable liquid at room or body temperature.[173] Very recently, Wolbert et al.[173] applied the UNIFAC(Dortmund) $g^E$-model to obtain activity coefficients for equation (4) in order to estimate the melting point depression of THEDES as a function of API concentration. The predictions obtained from the group contribution model are in very good agreement with the experimental data and thus could potentially reduce experimental efforts to identify adequate API/excipient eutectic systems.

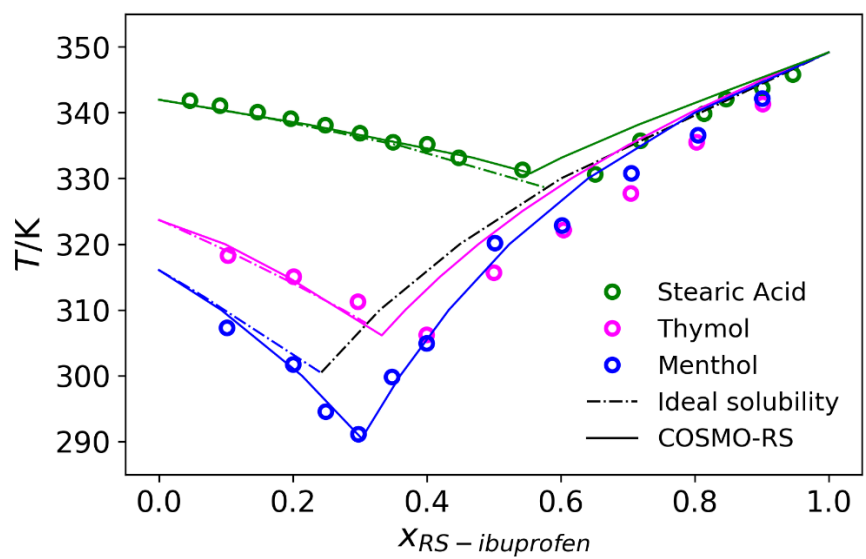

**Figure 5.** Solid-liquid equilibrium diagram of therapeutic eutectic building species. Experimental data (O, O) from Stott et al.[179] and (O) Corvis et al.[175] Dashed lines (-··-) represent ideal solubility as calculated in this work with equation (4) and values from Table 1. Solid lines (—) are COSMO-RS[16,180] predictions (COSMOthermX[181] BP_TZVP_C30_1701).

For the case of non-ionic DESs, COSMO-RS may prove a robust predictive tool for qualitative analysis as shown in Figure 5 for ibuprofen-based eutectic mixtures. This model is of aid particularly when a binary interaction parameter between chemical groups in UNIFAC is unknown and experimental data is unavailable. However, as described previously, COSMO-RS-based SLE modeling of DESs and eutectic systems containing ionic species can be challenging[2,167] and may require further model development.

Finally, it is common knowledge that the experimental measurements for eutectic point determinations may be strongly influenced by water[1,8,173,182] and care must be taken that this is done under anhydrous conditions. Additional considerations must be accounted for in special cases, particularly when handling subtances with many stereoisomers. For instance, two different melting points of menthol / lauric acid-based eutectic mixtures are described in the literature.[7,162,183] One contains L-menthol and the other racemic DL-menthol. Equilibrium with two ideal pure solids would no longer be a valid assumption in equation (4) at varying D- to L-menthol ratios if crytalline polymorphs[174] formed preferentially. This also opens questions regarding how certain isomer ratios could affect the deviations from ideal behavior as well as the recoverability of the eutectic mixture. In addition, there are up to 8 menthol isomers and various polymorphs[184] reported in the literature. Under the assumption that the data reported for eutectic point measurements in our review used the same D- and L-menthol isomers, then a hypothetical ternary system could be represented as something similar to Figure 6.

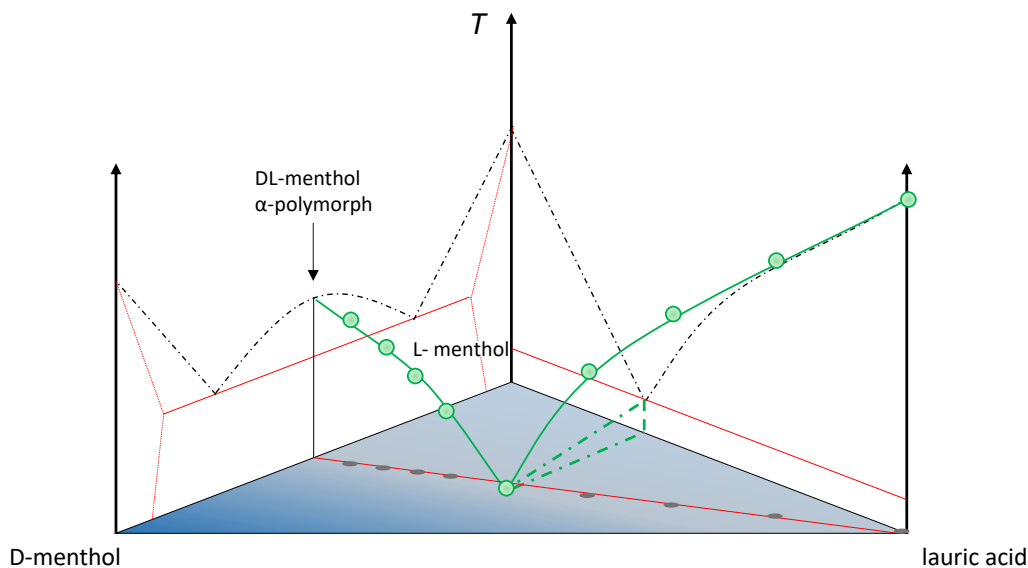

**Figure 6.** Hypothetical representation of a D-menthol, L-menthol, lauric acid ternary system qualitatively based on the data reported by Corvis et al.[174] (SLE for D- and L- menthol isomers, left pane), Martins et al.[7] (SLE for L-menthol/lauric acid, right pane) and Verma and Banerjee[162] (SLE for DL-menthol/lauric acid, green points).

Figure 6 is a qualitative hypothetical representation of what the ternary SLE of D-menthol, L-menthol and lauric acid could look like. The solid phase in the case of the DL-menthol mixture was not analyzed[162] but it seems reasonable to assume that the 1:1 liquid racemate follows the solid line of the most stable DL-crystalline polymorph and thus the mixture can be handled as binary. For varying isomer ratios, it may be the case that more than one polymorph and meta-stable polymorphs form in the mixture as the activity coefficients and concentrations of the isomers diverge. Therefore, actual experimental measurements at different isomer ratios might be challenging. In this regard, thermodynamic modeling could play a role in providing a qualitative/quantitative guess to shed light on such specific observations.

## 3.5. Solubilities

The solubility of a solid component in a pure DES or in a mixture containing DES forming components can also be modeled via equation (4) and either predict activity coefficients or fit an empirical model to the experimentally determined data. The latter is the most common method.

As mentioned in the previous section, one of the assumptions of equation (4) is that the system that is being described is in equilibrium with ideal solids ($x_i^s \gamma_i^s = 1$). This assumption may hold for the deep eutectic mixtures reported so far, but it may not necessarily be applicable for more complex multicomponent mixtures. It is common practice[25,185–187] to deal with the DES as an indivisible unit. It is well documented[8,138,188] that the effect of HBD compounds like water or alcohols may break the HBA-HBD interactions in a DES beyond a certain concentration, thus the mixture would be more of a multicomponent solution than a DES + n-components system.

For instance, the solubility of diverse sodium salts in ChCl-based DESs has been reported in the literature[185] and a non-electrolyte NRTL is fitted to the experimental data. Such a modeling approach assumes that the DES is an indivisible unit, a pure component, and that the sodium salt is also a pure component (a neutral quasi-dipolar ion pair). Whether the ions dissociate or not, and how ions affect the structure of the DES remain unanswered questions regarding non-local electrostatics. Many HBDs used in DES systems are alcohols and polyalcohols which have acceptor groups capable of solvating a sodium cation. In addition, when adding e.g. sodium bromide to a ChCl-based DESs, there is not one but at least four hypothetical solids: ChCl, choline bromide, sodium chloride, sodium bromide and potentially any hydrated form. While the neutral NRTL model fits to the measurements, it must be considered that this is a highly empirical model that can describe a wide variety of data without much physical constrains and thus, questionable insights may be obtained from the resulting parameters. Unfortunately, treating DES-containing systems as electrolytes is usually avoided.

Gano et al.[187] report the solubility of thiophene and dibenzothiophene in iron and zinc based type I DESs (a DES with a transition or post-transition metal chloride as anion[6] e.g. imidazolium + chloroaluminate) and fitted an electroneutral NRTL model to the experimental data. Mokhtarpour et al.[25] report the solubility of naproxen and compared the quality of the fit of e-NRTL, Wilson and UNIQUAC based models with an electrolyte term. While the selection of the modeling strategies is not thoroughly discussed and the models correlate to the experimental data, a clear example of why a DES should not always be treated as an indivisible unit in presence of a third component is described by Domańska et al.[186] In their work, the {ChCl+phenylacetic acid (1:1)} (DES) + water and the {ChCl+phenylacetic acid (1:2)} (DES) + water systems are reported. The experimentally determined eutectic point of the mixtures are shown in an SLE diagram where the abscissa is the mole fraction of the DES and the ordinate is the temperature. The eutectic point lies above the melting point of water in what is reported as an SLE + LLE system, that is, actually an SLE system with a miscibility gap. This could indicate that the equilibrium in the mixture involves some other hypothetical solid(s) that may contain DES constituents. As a consequence, treating the DES as an indivisible HBA:nHBD complex, particularly at high water concentrations (see Section 6.1), is not a good assumption.[8,188] Furthermore, if the eutectic nature of the DES is not analyzed then it is uncertain whether there are strong negative deviations from ideal behavior and consequently, a strong HBA-HBD interaction. This means the pseudo-component approach could be flawed if the strong HBA-HBD interaction is assumed but not confirmed.

Finally, a thorough computational methodology is provided by Jeliński and Cysewski[189] who applied a COSMO-RS based approach to correlate the solubility and infinite dilution activity coefficients of rutin in natural deep eutectic solvents (NADES). The authors found that applying the associated forms of the NADES components leads to a poor correlation. Subsequently, the dominant ionic forms and their concentration in the NADES were determined via chemical reaction equilibrium considerations. After this modification, a successful correlation between the solubility and infinite dilution activity coefficients of rutin in the different NADES was obtained and it could be satisfactorily applied for solvent screening.[189]

Two general conclusions can be drawn from the SLE calculations reported for DESs in the literature. Firstly, there is a clear need for more fusion properties of DES building components in order to tell deep from simple eutectic systems. Arbitrarily assuming that a system containing any typical HBA + any typical HBD will form a *deep eutectic* mixture without any additional consideration may lead to erroneous solvent screening and potentially unnecessary experimental efforts. Secondly, while the pseudo-component approach has proven to be satisfactory, the idea that DES are deep eutectic ***solvents*** in the literal sense of the word (an indivisible ion pair + HBD complex) might lead to misinterpretations. It excludes the fact that these are actually deep eutectic ***mixtures*** that will break up into their building components as soon as their energetic HBA-HBD interaction is not thermodynamically favorable. The addition of any component into the mixture plays a potentially crucial role in this last point.

4. Thermophysical Properties

This section reviews the modeling of different thermophysical properties of DESs. If available, the literature is discussed in the following order: $g^E$-models, EOS, MD simulations. Additionally, a table with an overview of all modelled systems and corresponding properties is provided in the supporting information.

## 4.1. Density

EOS can be used to model liquid densities of DES systems, but the main goal of the modeling is usually not the density per se. As shown in section 3.2 ($\varphi$ - $\varphi$ concept), the densities are employed to parametrize the EOS in some cases and apply it to the calculation of phase equilibria. For this reason, EOS are not mentioned here and the reader is referred to the above-mentioned section.

For MD simulations, the liquid densities are usually not used for fitting force field parameters, but they often deliver an initial estimation of the force field performance. By averaging the simulation box density over time, the liquid density as well as its dependency on the pressure and temperature can be studied in MD simulations. The use of reduced ionic charges was found to be necessary in the development of MD force fields. Scaling the charges in the mixture weakens the molecular interactions between the ions, resulting in an expansion of the system and therefore lower densities. This effect resulted in some cases in a better description of the density,[59,60,62] and in other cases however, in an underestimation.[68] Compared to the influence of many other force field parameters, the charge scaling was found to be the key factor.[59] Using scaled charges allows in general a proper description of the densities of many DESs at ambient conditions throughout different force field parametrizations. The best accuracy for a variety of DESs could be achieved by Doherty and Acevedo[78] with a mean error of 1.3% for several DESs over a temperature range of 298.15-333.15K. Besides Doherty and Acevedo, other studies[59,60,68,190] reported similar accuracy for the temperature dependent density, or more specific the thermal expansion coefficients. However, in case of the pressure dependency, the isothermal compressibility, an overestimation of 14.9-17.2% (or 0.41-0.98 $10^{-5}$ bar$^{-1}$, respectively) for the force field with the best overall performance was observed Ferreira et al.[68]. In their force field study, the parameter assignation, which predicted the density at ambient conditions accurately, also showed the best agreement with experimental pressure dependent densities but less accuracy for dynamic properties.

In addition to pure DES densities, the force fields GAFF[85] and MMFF [190] have been used to predict the densities of binary DES mixtures with water. Whereas a comparison with experimental densities of glycerol and water yielded a slight overestimation in the case of GAFF with charges of $\pm 0.8e$[85], Shah and Mjalli[190] reported an underestimation of 5-10 kg/m$^3$ using MMFF and charges of $\pm 1e$. This comparison shows the influence of the charge scaling on the resulting liquid phase densities, even in mixtures with a high water content.

### 4.2. Heat Capacities

Among the first developments of force fields suitable for DESs was their ability to predict the heat capacity. Perkins et al.[59] found a close agreement for the heat capacities of ChCl + urea, as the simulated properties lie within the experimental errors. Similar accuracy could later be achieved for DESs composed of ChCl and ethylene glycol[60] as well as glycerol.[60,62] When extending the applicability to a wider variety of DESs the heat capacities can be calculated with an overall accuracy of 5% using the force field parametrization of Doherty and Acevedo.[78]

### 4.3. Interfacial tension

Lloret et al.[130] combined soft-SAFT with density gradient theory to calculate interfacial tensions of quaternary ammonium chloride-based DESs. The results exhibit a correct trend and are applicable to quantitative evaluations after optimizing one parameter. Ideally this theory should be applied to a larger number of systems at different temperatures and pressures to estimate the influence of the modelled density by the EOS on the overall calculation.

Computing surface tensions from an NVT ensemble at room temperature resulted in significant overestimations in a prior study for ILs.[191] Similar issues have been reported by Ferreira et al.[68] for DESs with a mean model accuracy ranging from 1-10 mN m$^{-1}$ depending on the tested parameter set. The best performance was accompanied with force field parameters originally intended for ILs. The usage of polarizable force fields has shown to improve the calculation of surface tensions for ILs[58], thus similar studies for DESs seem to be promising. Doherty and Acevedo[78], on the other hand, computed the surface tensions solely at 425 K due to slow dynamics of the tested DESs. Good agreement with a mean error of 1.5 % (±0.65 mN m$^{-1}$) was achieved with experimental correlations, which extrapolated measured surface tensions from room temperature to 425 K using an Othmer equation.[192]

5. Dynamic Properties

Models regarding the viscosity and self-diffusion coefficient of DES systems are reviewed in this chapter. Combining EOS with a friction theory or free volume theory allows the computation of viscosities. On the other side, MD simulations deliver a more detailed picture of the dynamic behavior, as the movement and positions of all simulated molecules are studied over time. The difficulties in calculating dynamic properties of DESs may be related to the challenges reported for MD simulations for ILs. For example, the transport properties are underestimated by 50% for ILs in the simulations of Liu et al.[42] using GAFF parameters and ionic charges of $\pm 0.8e$.

5.1. Viscosity

Experimental and computational studies reported high viscosities for many DESs in comparison to other organic solvents becoming the main restriction of their application to for example bioactive systems.[193–195]

Regarding the calculation of viscosities from EOS, Haghbakhsh et al.[196] combined friction theory with the CPA-EOS and the PC-SAFT EOS to correlate viscosities of 27 different DESs at different temperatures and pressures. For this the different contributions of the EOS have to be divided into an attractive and a repulsive pressure term which then can be applied within the friction theory framework. Although the dipole moment and associating bond effect parameters were set fixed at the values of alcohols, the model was able to correlate all viscosity data with an average relative deviation of 4.4% for both EOS. However, because the friction theory requires 5 parameters adjusted for every DES the best way to estimate the applicability of this joint theory is to look at data that spans over large temperature and pressure ranges. In the set of 27 DESs only 3 include data at pressures other than at atmospheric conditions. For these systems indeed the average relative deviation spans from 0.7% to 18.3%. This includes the system with the highest deviation of the complete set. A revised version with only 3 parameters for the friction theory was published by same the group[197] reducing the average relative deviation for both EOS to 2.7%. Lloret et al.[130] combined soft-SAFT with free volume theory and further reduced the number of fitted parameters. Although the combined model was able to reproduce the viscosities for several quaternary ammonium chloride based DESs, all the data discussed was at atmospheric pressure. In general terms, it is possible to combine EOS with other theories to calculate viscosities. However, the relevance of the EOS input for the calculations requires further investigation.

The large viscosities of many DESs are often related to an extensive hydrogen bond network between the IL and the HBD[198] as well as in some cases to strong HBD-HBD interactions[199] (see section 6.1). This complicates their accurate prediction with MD simulations[68,74], as a proper sampling of the viscosity cannot always be reached due to slow dynamic behavior. The challenges start with the equilibration of the systems in MD simulations being affected by the slow dynamics of many DESs. This can be particularly said for the case of carboxylic acids as HBDs (e.g., malonic acid[60], levulinic acid[62]) where long simulation times are necessary for the equilibration of such viscous systems. Based on the equilibration of polymeric systems[200], Perkins et al.[59,60] suggested the usage of a compression and decompression scheme to ensure an efficient equilibration of high viscous DES systems, which was later adopted by Mainberger et al.[62]. After a successful equilibration, dynamic viscosities can be determined by applying the non-equilibrium periodic perturbation method[201,202], which relates the viscosity to the responding velocity profiles after an initial acceleration. However, the reliability of these results has to be checked carefully due to a strong influence of the applied initial acceleration on the sampled viscosities. In the framework of equilibrium molecular dynamics (EMD) viscosities can be estimated from a Green-Kubo expression[201,203,204], which relates the pressure tensor at each time step to its initial value. For that purpose, Maginn et al.[204] suggest to perform replicate simulations with varying initiations of the pseudo-random number generators in order to balance the large fluctuations of the autocorrelation function, which occur at long simulation times. Although both methods make it possible to obtain viscosities from MD simulations, convergence issues have been reported in particular for high viscous DESs at low temperatures.[85]

Concerning the prediction of viscosities, limited MD studies have been published in the past years. By utilizing the Green-Kubo relation García et al.[134] found an overestimation of the viscosity by 27-36% for their refined force field for ChCl+levulinic acid. Given the difficulties of calculating viscosities in MD simulations and considering the divergences with the reported experimental values, the authors stated this to be in reasonable agreement with experimental findings. The force field screening of Ferreira et al.[68], on the other hand, reported an underestimation of the experimental viscosities by a factor of 2.3-2.5 for the force fields performing the best for static thermodynamic quantities, such as the density or heat capacity. However, the trend of the viscosities over temperature is described properly for all the tested force fields. The accurate description of the strong temperature dependency of the viscosity from MD simulations was later confirmed by Baz et al.[85] for ChCl+glycerol using the GAFF force field. They additionally reproduced the large dependency of the viscosity on the water content resulting in a fair agreement of 18% with experiments at 360.15K. However, in case of lower temperatures less agreement with the experimental viscosities was related to the difficulties of converging MD simulations for high viscous solvents. The overall best agreement with experimental measured viscosities could be found for the OPLS-DES force field of Doherty and Acevedo.[78] Their refinement of the OPLS-AA force field delivered an excellent agreement with experimental findings with an average error of 1.6% (absolute deviations ranging from 0.2 - 6 mPa · s) for a variety of pure DESs over a temperature range of 298.15K to 348.15K.

In general, reducing the ionic charges in the IL allows a better description of the viscosities of many DESs, even though in some cases large deviations from the experimental observations still remain. This highlights the necessity of exploring force fields which include the molecular polarizability to better reflect the dynamic behavior of DESs.

### 5.2. Self-Diffusion Coefficient

In analogy to the viscosities, the prediction of self-diffusion coefficients of DESs with MD simulations is reported to be a challenging task due to their slow dynamics. In general, following the position of every molecule over the simulation time allows the calculation of mean squared displacements $\Delta r^2$ (MSD). The MSD can afterwards be related to the self-diffusion coefficient using the Einstein equation[205] and corrected for finite size effects.[85,206] It must be noted, that the sampling of the MSD should only be performed if the studied system is in the diffusive regime indicated by the parameter $\beta(t) = d\log(\text{MSD})/d\log(t)$ equal to unity. In several studies of DESs[59,68,78,85], this diffusive regime could not be reached within the simulation time in particular at low temperatures (280K-320K). This restricts the capability of accurately predicting DESs diffusion by MD simulations to elevated temperatures.

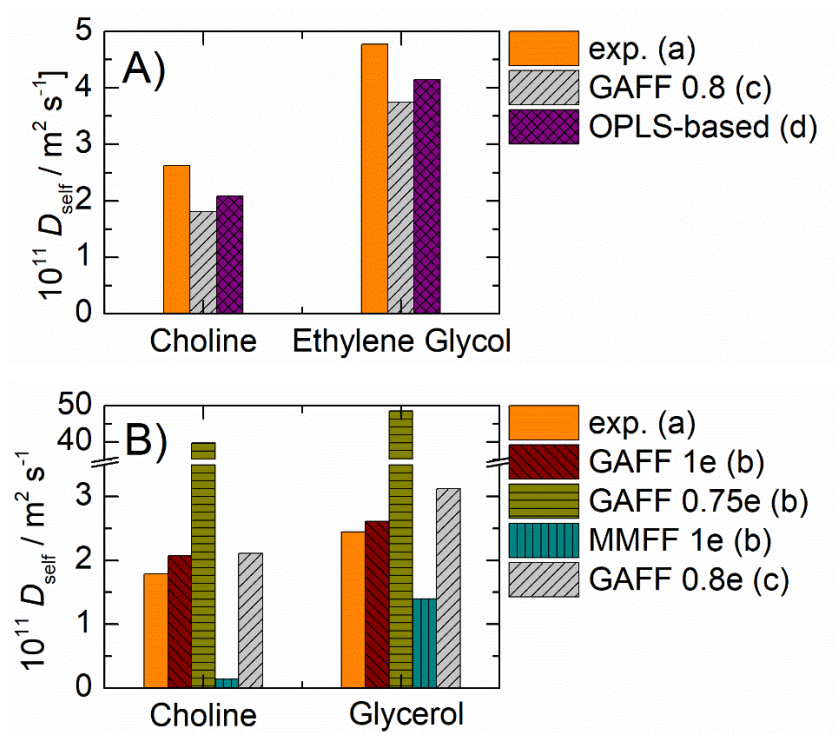

**Figure 7.** Comparison of self-diffusion coefficients for the eutectic mixture of (A) ChCl + ethylene glycol at 298 K and (B) ChCl + glycerol at 328 K from experiment and different MD simulations. (a) D'Agostino et al.[207], (b) Mainberger et al.[62], (c) Perkins et al.[60], (d) Ferreira et al.[68].

The calculation of self-diffusion coefficients has been discussed since the first development of force field models for DESs. Starting with the work of Perkins et al.[59,60], systematic underestimations by 15-50% of the self-diffusion coefficients at 298 K were observed for several ChCl- based DESs, for example ChCl + ethylene glycol (see Figure 7A). On the other hand, at 330 K the self-diffusivities were in good agreement for ChCl + urea and overestimated by 25% (0.22 $10^{-11}$ $m^2$ $s^{-1}$) for other DESs, for example ChCl+glycerol (see Figure 7B). Higher temperatures correspond to a reduction of solvent viscosity, thus allowing a proper sampling of the dynamic behavior. In addition, Perkins et al.[60] reported a large sensitivity of the charge scaling towards the predicted self-diffusion coefficients. For the implementation of ionic charges of $\pm 0.8e$ the self-diffusion coefficients are mainly overestimated, whereas GAFF systematically underestimated the self-diffusivities for ionic charges of $\pm 0.9e$. Based on a different parametrization of GAFF[63], Mainberger et al.[62] reported similar trends for their force field as shown in Figure 7B. When the ionic charges are reduced the predicted self-diffusion coefficients increase significantly and are overestimated by one order of magnitude in case of $\pm 0.75e$. Although the MMFF[67] performed better for predicting dynamic properties compared to GAFF without any charge scaling, the large difference of the self-diffusivities for the cation and HBD contradicts the results of experimental measurements and other MD simulations. By focusing on one specific DES, ChCl+ethylene glycol, and using scaled charges of $\pm 0.8e$, Ferreira et al.[68] could improve the results of Perkins et al.[59,60] when combining OPLS-AA force field parameters originally intended for ionic liquids[77] with parameters of Szefczyk and Cordeiro[74] for ethylene glycol (see Figure 7A).

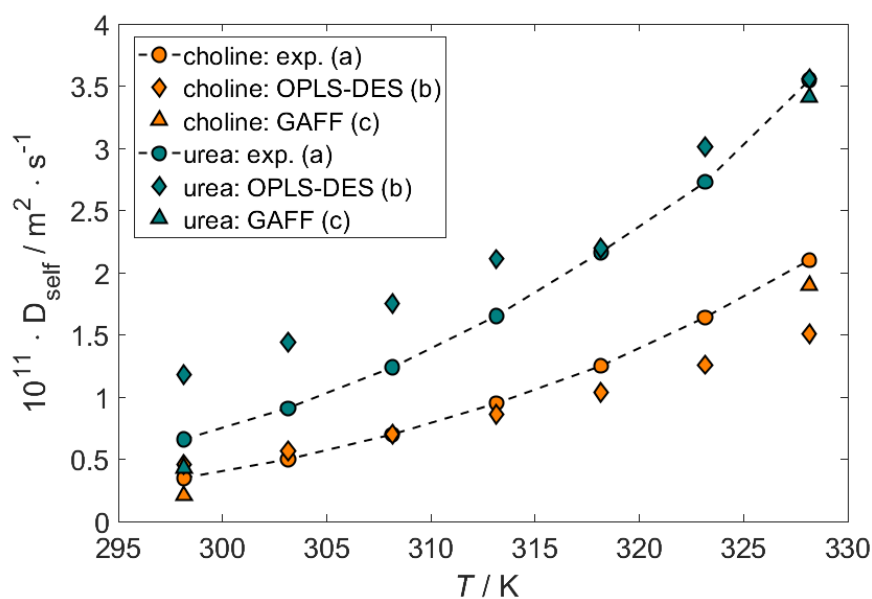

**Figure 8.** Comparison of self-diffusion coefficients for the DES ChCl + urea from different MD simulations. The dashed lines are included to guide the readers view. (a) D'Agostino et al.[207], (b) Doherty and Acevedo[78], (c) Perkins et al.[60].

In agreement with previous findings, Doherty and Acevedo[78] concluded, that a temperature above 400 K is necessary for an adequate sampling of the self-diffusivities of ChCl-based DESs. The diffusive regime could not be reached for every solvent at moderate temperatures of 298 - 328 K. However, they have extrapolated the self-diffusion coefficients of ChCl+urea, which were determined for temperatures above 400 K, to ambient temperatures by fitting $\ln(D_{\text{self}})$ versus $1000/T$. Figure 8 shows the resulting self-diffusion coefficients for choline and urea from Doherty and Acevedo[78] compared to experimental measurements of D'Agostino et al.[207] and the simulations of Perkins et al.[60] The temperature dependency of the self-diffusion coefficients in ChCl+glycerol cannot be reproduced well by the extrapolation of Doherty and Acevedo.[78] Whereby, the trend of the urea diffusivities supports the statement of an improved performance with increasing temperature. On the other hand, the choline trend contradicts this by having the closest agreement at 308 K. The reported values by Perkins et al.[60] show a better agreement with the experimental measurements for both species. However, the self-diffusion coefficients at different temperatures are slightly underestimated.

Besides the prediction of the dynamic properties of pure DESs, few studies have addressed the effect of water on the dynamic behavior of DESs. For example, Shah and Mjalli[190] utilized MMFF and found a significant increase the diffusion of the DES components above a water concentration of 25 wt%. They concluded that the strong anion-HBD interactions are replaced by interactions with water and therefore the mobility of the ions and the HBD rises with increasing water content.

This behavior was later confirmed by Baz et al.[85] in a study of calculating diffusion coefficients of water mixed with ChCl+glycerol for temperatures in the range of 280.15-360.15K.

In summary, due to high viscosities at low temperatures it is difficult to calculate diffusivities from MD simulations. Furthermore, the results of the force field parametrizations of various groups suggest that non-polarizable force fields are not well suitable of reproducing the self-diffusion coefficients of DESs at room temperature. The charge scaling in non-polarizable force fields give the best representation of other static thermodynamic properties like the density, but does not necessarily predict the dynamic behavior correctly. Treating the observed charge transfer in the DES with an average ionic charge, for example $\pm 0.8e$, does allow an adequate prediction of static properties. However, the real polarization state of the molecules may differ significantly from the average assumption and may affect the movement and interactions of the particles. To reasonably reproduce the dynamic behavior of ILs and DESs, these polarizabilities should possibly be considered. Therefore, Doherty and Acevedo[78] suggested the development of a polarizable force field to better describe the dynamic behavior of DESs. In case of ILs for example, the use of polarizable force fields essentially enhanced the quality of the prediction of diffusion coefficients for several ionic components.[58] Comparable improvements can probably be expected for simulations of DESs in the near future.

6. Further MD Studies

6.1. Liquid Structure

Few experimental investigations of the molecular interactions in DES mixtures have been performed. Therefore, it is difficult to experimentally determine the interactions between different molecules present in DESs. Examples of experimental measurement are pulsed field gradient nuclear magnetic resonance (NMR)[207] or neutron diffraction (ND) measurements.[208] Additionally to calculating their bulk phase properties, MD simulations are capable of reproducing the microstructure of a DES as described by radial distribution functions (RDF) and analyzing the hydrogen bond network. Therefore, MD simulations provide a suitable tool for understanding the interactions present in an eutectic mixture.

A recent review about the interactions present in deep eutectic mixtures was presented by Wagle et al.[209], who reported a disruption of the long-range order of the ionic compounds initiated by the HBD in many computational studies. This leads to reduced interaction energies within the deep eutectic mixture, possibly explaining the liquid state of a DES at moderate conditions. For instance, Perkins et al.[59] revealed a stronger hydrogen bond formed between urea and chloride, in particular the NH-hydrogen trans to the C=O bond, compared to the interactions of urea with the choline ion. These results are in agreement with experimental findings and have been confirmed, amongst others[190,210], by Sun et al.[211] by additionally observing a lowering of the interionic interactions and a disruption of the long-range ordering of choline and chloride. In a follow up study covering more DESs, Perkins et al.[60] observed, that the preferred interactions with the chloride ion are not unique to urea, but also present for other HBDs (e.g., glycerol, ethylene glycol). Similar patterns are also noticed for other types of DESs[212] and hint that this particular interaction between a HBD and the anion is responsible for the formation of the deep eutectics. However, Ashworth et al.[213] stated that hydrogen bonds between urea and choline are competitive to urea-chloride interactions and should be considered when modeling the formation of the ChCl+urea DES.

Water is found to have a significant impact on the hydrogen bond network of the DES. By adding water to the solution Shah and Mjalli[190] observed a replacement of the hydrogen bonds between urea-urea and urea-chloride by interactions with water, until finally at high water concentrations every DES component is individually hydrated resulting in low interactions and a higher mobility of the DES components. However, at low water concentrations a slight increase in the urea-urea hydrogen bonds could be observed.

Complementary to the review published by Wagle et al.[209], additional studies should be mentioned in this regard. An analysis of RDFs at different non eutectic mixtures by Mainberger et al.[62] delivered no significant deviation compared to the eutectic point because the probability for HBD-HBD interactions rises with increasing HBD content, as expected. They concluded, that the structure analysis from calculated RDFs is insufficient to properly predict the eutectic mixture with MD simulations. Mainberger et al.[62] found no clear dependency of any interactions with varying composition, thus suggesting that polarizable force fields can help to better understand and reproduce the formation of DESs. It has also to be mentioned that the charge scaling may properly reproduce the collective behavior of DESs. However, prior investigations of ILs have shown, that local interactions such as hydrogen bonds are underestimated by reducing the ionic charges.[58]

Concerning binary mixtures with water, Ahmadi et al.[214] observed no change in the glycerol-glycerol hydrogen bonds at a water mole fraction of 0.9. On the other side, the disruption of the DES hydrogen bond network by water has been shown for other DESs by Baz et al.[85] and Zhekenov et al.[215] in close agreement with the work of Shah and Mjalli[190]. For example, the hydrogen bond analysis of (ChCl+glycerol)-water mixtures by Baz et al.[85] showed that with increasing water concentration, the hydrogen bonds between the glycerol molecules disappear due to a preferred interaction of glycerol with water (see figure 4 of Baz et al.[85]). Similar effects were observed for the choline-glycerol hydrogen bonds, which are replaced by hydrogen bonds between choline and water, resulting in a vanishing of the DES characteristics at increasing water concentrations. Zhekenov et al.[215] concluded that for low water concentrations the hydrogen bond network of the DES remains intact resulting in little changes of the macromolecular properties.

However, for water mole fractions larger than 0.5 properties such as the diffusivities would significantly change. In spite of the de-structuring role of water, Weng and Toner[195] found, that water can also replace the chloride ions in terms of bridging choline and glycerol. With having its maximal impact at 35.8 wt% of water, this may give guidance in optimizing the water content in a DES for specific application such as bio-catalysis, without destroying the hydrogen bond characteristics of the DES. In agreement with prior studies as well as experimental ND measurements combined with an empirical potential structure refinement model[208] and first principle MD[216], the refined OPLS-DES force field by Doherty and Acevedo[78] correctly reflects the interactions of the HBD and the chloride ion to be the predominant ones in the corresponding DES. Complementary to ChCl-based DESs, Kaur et al.[217] found a similar spatial heterogeneous structure in other DESs composed of $LiClO_4$ and alkylamide, which is caused by ion-pair self-segregation and is enhanced at elevated temperatures.[218]

In summary, MD simulations deliver useful insights into the interactions present in deep eutectic mixtures. As reported by many studies, the interactions between the HBD and the anion are supposed to be the predominant ones in DESs and are probably able to explain their liquid state at ambient conditions. However, the formation of DESs cannot be predicted by the structural analysis of the results from non-polarizable MD simulations, demanding future research on polarizable force fields.

A thorough review about the influence of water on the structure of ILs and DESs studied with MD and Monte Carlo (MC) simulations and with advanced experimental methods can be found in the work of Ma et al.[13]

6.2. Examples of DES Applications Studied with MD

Besides elucidating the intermolecular interactions (Section 6.1) present in a DES as well as calculating macroscopic properties (sections 3, 4 and 5), another scope of MD simulations is the investigation of the solvent behavior for particular applications. Whereby, MD simulations offer diverse application possibilities and can for example be used to study the gas capture in DES (e.g., fuel desulfurization[219], mercury capture[220]). Besides that, the behavior of DESs on the surface of metallic components[221], in particular electrodes[222], as well as in the presence of nano-structures[223,224] can be elucidated. In the past years, the effect of external electrostatic fields[222,225] on DES molecules as well as exploring the suitability of DESs for batteries[226] became popular. It has to be mentioned, that this overview is of exemplary nature in order to demonstrate the various opportunities of applications studied with MD. As a growing field of research, the use of DESs has additionally been attracting great attention in the context of finding sustainable solvents for bio-catalytic systems.[87,227–231] Regarding tailored characteristics, DESs offer great opportunities in tuning the solvent effects in bio-catalysis. Furthermore, few studies of DNA structures in DESs and ILs have been published.[232–234]

Besides several experimental investigations of peptide systems, only two MD simulation studies with enzymes and DNA sections in a DES have been published in the past years. Monhemi et al.[227] observed that the enzyme Candida Antarctica Lipase B (CALB) can be stable in solution containing high concentrations of urea if it is present as a DES in combination with ChCl. An in-depth analysis using MD simulations observed complexes of choline and urea, which are formed at the proteins surface and present the denaturation of CALB by urea. The immobilization of urea at the surface of the enzyme also maintains the functionality of the active site and the catalytic activity. Pal and Paul[234] analyzed the effect of (ChCl+urea)-water mixtures on guanine-rich quadruplex Thrombin Binding Aptamer (TBA) using MD simulations. A deviation of the structure of TBA in MD simulations from experimental NMR measurements as well as an increasing rigidity of the DNA in low-water solutions could be explained by favored hydrogen bond interactions of the DNA with urea.

However, Pal and Paul[234] also questioned whether the parametrization of GAFF[59,60] can accurately represent the DNA-solvent interactions due lacking experimental validation. This is a general problem of MD force fields that is difficult to be resolved. However, having carefully checked the reliability of the used force fields, they offer a great ability to explore and elucidate the DES behavior observed in experiments.

7. Conclusions

A lot of efforts have been made in the recent years to model DES systems. However, most of the reviewed literature focusses on modeling the same class of systems: quaternary ammonium chlorides and bromides and ChCl as HBAs using a wide variety of different species as HBDs. The approaches reviewed in this paper are summarized in Table 2 accompanied with the main observations /peculiarities concerning each of them

**Table 2.** Key observations and conclusions regarding modeling different properties of DESs.

| Property | Modelling approach | Key conclusions |
|---|---|---|
| Activity coefficient | EOS | Application of the individual constituent approach seems to deliver more thermodynamically consistent results and is more transferable to other systems than the pseudo-component approach. |
| | MD | Thermodynamic integration is computationally demanding compared to EOS and gE-models.<br><br>Advantageous for studying $\gamma_i$ in the presence of a macromolecule (e.g., enzyme). |
| | | Large sensitivity towards the free energy difference calculated from MD simulations. |
| VLE | $g^E$-model & EOS | Assuming that the DES has a low vapor pressure and neglecting its effect seems to be a good assumption in most cases. |
| | | The relevance of using a model to calculate the non-idealities of these systems has not been thoroughly discussed in the literature. As with other systems, this relevance increases with increasing pressure and decreases with increasing temperature. |
| | | A lot of modelling approaches employ a combination of gE-model and EOS, whereas the experimental data suggests that, at least for the systems studied so far, the assumptions of Henry's law are applicable. |
| | | The modified Lyderson-Joback-Reid group contribution method in combination with the Lee Kesler mixing rules allow estimating the needed critical properties and acentric factor of DESs with different HBD to HBA ratios. However, when applying cubic EOS binary interaction parameters are usually needed to correlate the data quantitatively. |
| | | SAFT based EOS can be successfully applied by adjusting the parameters to liquid density data. |
| | | Employing the individual constituent approach delivers models that are more applicable to different conditions and HBD to HBA ratios even allowing prediction of VLE without the need of adjusting binary interaction parameters. |
| LLE | $g^E$-model & EOS | The non-electrolyte NRTL provides a good fit when using the pseudo-component approach, provided that this is done in a system where there is a DES rich phase and little to no DES present in the other phase. |
| | | The pseudo-component approach may still be valid when the DES concentration in both phases is significant only if the integrity of the DES in both phases is confirmed. |
| | | COSMO-RS based approaches may use the individual constituent or the pseudo-component approach. However, when the concentration of the DES is significant in both phases the individual constituent approach is more suitable. |
| | | The individual constituent approach in COSMO-RS is considered to be more consistent when the salt (in ionic DESs) |
| | | is modeled with the electroneutral approach instead of the ion pair approach. |
| | | The PC-SAFT equation of state has proven effective when correlating LLE with the pseudo-component approach. Accurate results nevertheless require the adjustment of binary interaction parameters. |
| SLE (Eutectic Point) | $g^E$-model & EOS | The solid-liquid equilibrium diagrams are central to determining whether a eutectic system can be classified or not as a DES based on the presence of strong negative deviations form ideal behavior. |
| | | PC-SAFT has been commonly and successfully applied for the correlation of eutectic points for ionic and non-ionic DES forming mixtures. |
| | | COSMO-RS can be applied as a predictive model for non-ionic DES forming mixtures but cannot be applied to ionic systems unless its energy interaction equations are trained with experimental data. |
| | | If other $g^E$-models are applied to correlate experimental data, a modified Redlich-Kister expansion[2] seems to outperform the NRTL model. |
| | | UNIFAC(Dortmund) has been successfully applied for the prediction of eutectic points of non-ionic DES-forming therapeutic mixtures. |
| | | SLE calculations can only be performed if the fusion data of the components is known. For various DES forming compounds these may have to be estimated, as the decomposition temperatures result in biased experimental values. |
| Solubility | $g^E$-model & EOS | The solubility of a component in a DES is commonly determined by experiment and a model like NRTL is correlated to the data applying the pseudo-component approach. This approach may prove inaccurate as the number of components in the mixture rises, particularly if these are of ionic nature. |
| | | A COSMO-RS based methodology that correlates infinite dilution activity coefficients with experimental solubility values has been described in the literature.[189] |
| Density | MD | Good agreement with experimental density of pure DESs as well as in mixtures with water for a variety of force fields. |

| | | Thermal expansion coefficients correctly reflected by several force fields |
| --- | --- | --- |
| | | Force field models tuned for the isothermal compressibility fail to predict the dynamic behavior. |
| Heat capacity | MD | Agreement with experimental measurements within an error of 5%. |
| | EOS | Combination with density gradient theory allows the correlation of interfacial tensions, however due to the narrow application range it is questionable how transferable these models are to other conditions. |
| Interfacial tension | MD | Overestimated similar to prior IL studies. |
| | | Challenging due to sluggish dynamics of many DESs. |
| | | Good agreement with experimental observations if calculated at elevated temperatures followed by an extrapolation to ambient conditions. |
| | EOS | Combination with friction theory or free volume theory allows the correlation of viscosities. However, due to the narrow application range it is questionable how transferable these models are to other conditions. |
| Viscosity | MD | Ionic charge scaling improves the viscosity prediction. |
| | | Convergence issues at ambient temperatures due to slow dynamics. |
| | | Inconsistencies between different force field models. |
| | | Force fields with best performance for static properties do not necessarily reproduce the viscosities correctly. |
| | | Best agreement found for the OPLS-DES force field.[78] |
| Self-diffusion coefficients | MD | Convergence issues related to the slow dynamics. |
| | | Charge scaling significantly effects self-diffusion coefficients with inconsistencies reported for different force fields. |
| | | Deviations between simulated and experimental values up to one order of magnitude reported. |
| | | Doherty and Acevedo[78] suggested the calculation of diffusion coefficients at elevated temperatures followed by an extrapolation to ambient conditions. |

|  |  | The observed challenges suggest that polarizabilities should be included in the force field models in order to better reproduce the dynamics of DESs. |
| --- | --- | --- |
| Liquid structure | MD | Disruption of long-range order of the ionic components initiated by the HBD. |
|  |  | Strong interactions between HBD and anion are of particular importance in understanding the formation of DESs. |
|  |  | Polarizable force fields are suggested to further guide understanding the formation of DESs. |
|  |  | Hydrogen bonds may be underestimated due to the reduced ionic charges in non-polarizable force fields. |
|  |  | Disrupting role of water on the hydrogen bond network resulting in an individual hydration of the DES components as water concentration rises. |

$g^E$-models and EOS have been successfully applied to calculate VLE, LLE, SLE and solubilities of other components in a DES containing system. For VLE it seems that in some cases the models employed could be simplified.

The pseudo-component approach (modeling the DES as a complex) is widely used due to its simplicity, while the individual constituent approach is only applied in fewer cases. Although UNIFAC[235], COSMO-RS[18,180] and a few EOS are applied predictively for some systems, in the majority of the literature the models are used to correlate experimental data. In the future it would be beneficial to concentrate more efforts on the individual constituent approach as it would lead to more predictive and transferable models applicable for systems for which a small amount or no data is available. In this regard, group contribution methods like GC-PC-SAFT[236,237], SAFT-$\gamma$ Mie[238], UNIFAC[235] and other models like COSMO-RS[18,180] can be promising. For IL/salt based DESs maybe electrolyte models like ePC-SAFT[239], AIOMFAC[240] or COSMO-RS-ES[38,241] might be applicable. These models would also speed up the screening for DESs. While $g^E$-models and EOSs are advantageous for the calculation of phase equilibria, MD simulations offer insights into the liquid structure as well as macroscopic thermophysical and dynamic properties.

The use of non-polarizable force fields and a scaling of the ionic charges, which reflects the charge transfer as well as polarization effects, is a promising alternative to computationally demanding polarizable force fields. While this technique has proven capable of representing static and collective properties as well as the liquid structure of many DESs, the prediction of dynamic properties has shown the limits of non-polarizable force fields and hint the necessity of developing advanced methods to increase the accuracy of dynamic property predictions. The diverging individual polarizations of every molecule may have a major impact on their movement and thus on the determination of dynamic properties from MD simulations. As suggested by several authors,[62,78] the development of polarizable force fields are recommended as a future field of research for the properties of DESs.

The application of MD simulations to analyze the interactions present in mixtures of DESs with macromolecules is a growing research topic. With the refinement of existing force fields originally developed for bio-molecular systems in order to handle DESs, these have become applicable for future MD studies of, for example, proteins in DESs.

Finally, more efforts are needed in defining when a mixture is or is not a DES. As Martins et al.[8] have pointed out, some of these systems behave quite ideally and a deep eutectic must be qualitatively differentiated from a simple eutectic before being arbitrarily described in the literature as a DES.

**Supporting Information**.

A table with the references, properties, substances and models that were applied in the reviewed literature in included in the supporting information.

**Corresponding Author**

*corresponding author (irina.smirnova@tuhh.de)

**Author Contributions**

The manuscript was written through contributions of all authors. All authors have given approval to the final version of the manuscript.

**Funding Sources**

The authors appreciate the financial support of the DFG (project-ID: 391127961).

**Acknowledgements**

The authors wish to express their gratitude to Professor João A.P. Coutinho from the University of Aveiro and Professor Tamal Banerjee from the Indian Institute of Technology Guwahati for their fruitful comments and discussion on the menthol / lauric acid eutectic systems.

(65) Zahn, S.; Kirchner, B.; Mollenhauer, D. Charge Spreading in Deep Eutectic Solvents. *ChemPhysChem* **2016**, *17*, 3354–3358.

(66) García, G.; Atilhan, M.; Aparicio, S. The Impact of Charges in Force Field Parameterization for Molecular Dynamics Simulations of Deep Eutectic Solvents. *J. Mol. Liq.* **2015**, *211*, 506–514.

(67) Zoete, V.; Cuendet, M. A.; Grosdidier, A.; Michielin, O. SwissParam: A Fast Force Field Generation Tool for Small Organic Molecules. *J. Comput. Chem.* **2011**, *32*, 2359–2368.

(68) Ferreira, E. S. C.; Voroshylova, I. V.; Pereira, C. M.; D. S. Cordeiro, M. N. Improved Force Field Model for the Deep Eutectic Solvent Ethaline: Reliable Physicochemical Properties. *J. Phys. Chem. B* **2016**, 120, 10124–10137.

(69) Ullah, R.; Atilhan, M.; Anaya, B.; Khraisheh, M.; García, G.; ElKhattat, A.; Tariq, M.; Aparicio, S. A Detailed Study of Cholinium Chloride and Levulinic Acid Deep Eutectic Solvent System for $CO_2$ Capture via Experimental and Molecular Simulation Approaches. *Phys. Chem. Chem. Phys.* **2015**, *17*, 20941–20960.

(70) Breneman, C. M.; Wiberg, K. B. Determining Atom-Centered Monopoles from Molecular Electrostatic Potentials. The Need for High Sampling Density in Formamide Conformational Analysis. *J. Comput. Chem.* **1990**, *11*, 361–373.

(71) Besler, B. H.; Merz, K. M.; Kollman, P. A. Atomic Charges Derived from Semiempirical Methods. *J. Comput. Chem.* **1990**, *11*, 431–439.

(208) Hammond, O. S.; Bowron, D. T.; Edler, K. J. Liquid Structure of the Choline Chloride-Urea Deep Eutectic Solvent (Reline) from Neutron Diffraction and Atomistic Modelling. *Green Chem.* **2016**, *18*, 2736–2744.

(209) Wagle, D. V.; Deakyne, C. A.; Baker, G. A. Quantum Chemical Insight into the Interactions and Thermodynamics Present in Choline Chloride Based Deep Eutectic Solvents. *J. Phys. Chem. B* **2016**, *120*, 6739–6746.

(210) Gao, Q.; Zhu, Y.; Ji, X.; Zhu, W.; Lu, L.; Lu, X. Effect of Water Concentration on the Microstructures of Choline Chloride/Urea (1:2) /Water Mixture. *Fluid Phase Equilib.* **2018**, *470*, 134–139.

(211) Sun, H.; Li, Y.; Wu, X.; Li, G. Theoretical Study on the Structures and Properties of Mixtures of Urea and Choline Chloride. *J. Mol. Model.* **2013**, *19*, 2433–2441.

(212) Kussainova, D.; Shah, D. Structure of Monoethanolamine Based Type III DESs: Insights from Molecular Dynamics Simulations. *Fluid Phase Equilib.* **2019**, *482*, 112–117.

(213) Ashworth, C. R.; Matthews, R. P.; Welton, T.; Hunt, P. A. Doubly Ionic Hydrogen Bond Interactions within the Choline Chloride–Urea Deep Eutectic Solvent. *Phys. Chem. Chem. Phys.* **2016**, *18*, 18145–18160.

(214) Ahmadi, R.; Hemmateenejad, B.; Safavi, A.; Shojaeifard, Z.; Shahsavar, A.; Mohajeri, A.; Heydari Dokoohaki, M.; Zolghadr, A. R. Deep Eutectic–Water Binary Solvent Associations Investigated by Vibrational Spectroscopy and Chemometrics. *Phys. Chem. Chem. Phys.* **2018**, *20*, 18463–18473.
89

**For Table of Contents use only.**

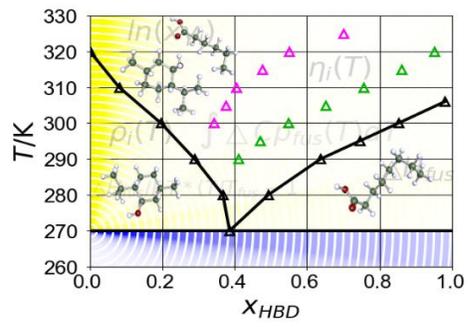